# Defect-Mediated Phase Engineering of 2D Ag at the Graphene/SiC Interface

Arpit Jain[1], Boyang Zheng[2,3], Sawani Datta[4], Kanchan Ulman[5], Jakob Henz[6], Matthew Wei-Jun Liu[1], Van Dong Pham[7], Wen He[5,8], Chengye Dong[1,3,9], Li-Syuan Lu[1], Alexander Vera[1], Nader Sawtarie[10], Wesley Auker[11], Ke Wang[11], Bob Hengstebeck[11], Zachary W. Henshaw[12], Shreya Mathela[13], Maxwell Wetherington[11], William H. Blades[12], Kenneth Knappenberger[13], Ursula Wurstbauer[6], Su Ying Quek[8,14,15], Ulrich Starke[4], Shengxi Huang[16], Vincent H. Crespi[2], and Joshua A. Robinson*[,1,8,17]

[1]Department of Materials Science and Engineering, The Pennsylvania State University, University Park, PA 16802, USA

[2]Department of Physics, The Pennsylvania State University, University Park, PA 16802, USA

[3]2-Dimensional Crystal Consortium, The Pennsylvania State University, University Park, PA 16802, USA

[4]Max-Planck-Institut für Festkörperforschung, Heisenbergstraße 1, 70569 Stuttgart, Germany

[5]Department of Physics, National University of Singapore, Singapore 117551

[6]Institute of Physics and Center for Soft Nanoscience (SoN), University of Münster, 48149 Münster, Germany

[7]Paul-Drude-Institut für Festkörperelektronik, Hausvogteiplatz 5-7, Leibniz-Institut im Forschungsverbund Berlin e. V., 10117 Berlin, Germany

[8]Department of Materials Science and Engineering, National University of Singapore, Singapore 117575, Singapore

[9]Center for 2-Dimensional and Layered Materials, The Pennsylvania State University, University Park, PA 16802, USA

[10]Department of Chemical and Petroleum Engineering, University of Pittsburgh, Pittsburgh, Pennsylvania 15260, USA

[11]Materials Research Institute, The Pennsylvania State University, University Park, PA 16802, USA

[12]Department of Physics and Engineering Physics, Juniata College, Huntingdon 16652, Pennsylvania, USA

[13]Department of Chemistry, The Pennsylvania State University, University Park, PA 16802, USA

[14]Centre for Advanced 2D Materials, National University of Singapore, Singapore 117542, Singapore

[15]NUS Graduate School, Integrative Sciences and Engineering Programme, National University of Singapore, Singapore 117456, Singapore

[16]Department of Electrical and Computer Engineering and the Rice Advanced Materials Institute, Rice University, Houston, TX, USA

[17]Center for Atomically Thin Multifunctional Coatings, The Pennsylvania State University, University Park, PA 16802, USA

[*] Corresponding author: jar403@psu.edu

**Abstract**

**Atomically thin silver (Ag) films offer unique opportunities in plasmonic, quantum optics, and energy harvesting, yet conventional growth methods struggle to achieve structural control at the monolayer limit. Here, we demonstrate phase-selective synthesis of large-area, crystalline 2D Ag films via defect-engineered confinement heteroepitaxy (CHet) at the epitaxial graphene/silicon carbide (EG/SiC) interface. By tuning graphene growth and post-growth defect introduction, two distinct Ag phases are achieved with disparate properties: a nearly commensurate $Ag_{(1)}$ lattice stabilized by vacancy and line defects in epitaxial graphene, and a denser $Ag_{(2)}$ phase preferentially grown with $sp^3$-rich zero-layer graphene. Structural and spectroscopic characterization confirm lattice registry with the SiC substrate, while theoretical calculations reveal a thermodynamic preference for $Ag_{(2)}$ but an easier nucleation for $Ag_{(1)}$. Both phases are found to be semiconducting, with the $Ag_{(2)}$ phase exhibiting slightly enhanced n-doping of graphene. Notably, nonlinear optical measurements reveal a three-order magnitude difference in second-order susceptibility between the two phases, demonstrating promise for phase-tunable 2D metals in reconfigurable optoelectronic and metamaterial platforms.**

## Introduction

Atomically thin, two-dimensional silver (Ag) exhibits exceptional optical and plasmonic properties,[1–3] with promise for bio-sensing, telecommunications, energy, and computation technologies. Key to exploiting the properties of silver nanofilms is the control over structural phases with different physicochemical properties.[4,5] However, contemporary evaporative[3,6,7] and sputtering[8–10] techniques result in insufficiently coalesced films in the monolayer and few-layer regime with limited structural phase control.

A promising alternative approach to 2D metal synthesis is intercalation, whereby metal atoms are sandwiched within layers of a host material. Intercalation of metals at the epitaxial graphene (EG) and silicon carbide (SiC) interface[11–13] has been a key route to modulating the graphene overlayer,[14] as well as stabilizing atomically thin allotropes of isotropic 3D metals,[15–20] including noble metals.[21–30] Furthermore, by intentionally introducing defects into EG prior to intercalation at near-ambient pressures, large-area coalescence of 2D metals can be readily achieved in a process dubbed confinement heteroepitaxy (CHet).[31] Remarkably, control over the process parameters during intercalation can result in different structural phases for the intercalant.[16,32–34] As a pertinent example, Forti *et al.* observed a phase transformation from a semiconducting monolayer Au with $(1 \times 1)$ registry over SiC(0001) substrate to a metallic $(20 \times 20)$ superstructure of bilayer Au atoms over a $(19 \times 19)$ SiC lattice, by changing the gold deposition time during growth.[24]

To date, monolayer and few-layer 2D-Ag have been synthesized on SiC both with[12,35] and without[25] plasma assistance. 2D-Ag can form both a triangular $(1 \times 1)$ lattice[25] with an indirect bandgap of ~1.1 eV[36] ($Ag_{(1)}$) and a denser phase with a unique spectral response in Raman spectroscopy ($Ag_{(2)}$).[37] In addition, a recent study has uncovered a large, phonon-based quantum interference signature within the EG/2D-Ag/SiC system which is modulated by the structural phase of 2D-Ag and the stacking order of SiC.[38] These property modifications further exemplify the need to control the 2D-Ag phase; however, direct control over phase formation has not been reported. Here, by tuning the pre-intercalation processing steps (graphene growth and defect engineering) in the CHet method, we achieved near-complete control over 2D-Ag phase selection. The denser $Ag_{(2)}$ preferentially forms under $sp^3$ defects in zero-layer graphene (ZLG or buffer layer), adopting a $(3\sqrt{3} \times 3\sqrt{3})R30°$ lattice over a (5×5) SiC unit cell confirmed through cross-sectional scanning tunneling electron microscopy (STEM), low-energy electron diffraction (LEED) and Auger electron spectroscopy (AES) mapping. Whereas the near commensurate $Ag_{(1)}$ phase preferentially forms under vacancy and line defects induced by helium plasma treatment in monolayer epitaxial graphene, although Rosenzweig *et al.* report intercalation of the $Ag_{(1)}$ phase under ZLG by thermal deposition in ultra-high vacuum.[25]

First-principles density functional theory (DFT) modelling indicates that the $Ag_{(2)}$ phase is more thermodynamically stable. It is confirmed experimentally as the $Ag_{(1)}$ phase partially converts to $Ag_{(2)}$ over time. However, the calculation also shows that $Ag_{(1)}$ intercalation is kinetically preferred due to better templating to the SiC lattice, leaving the opportunity to tune the relative areal coverage between these two phases during growth, which is realized experimentally in this paper. Angle-resolved photoelectron emission spectroscopy (ARPES) shows that both Ag-SiC phases represent a semiconducting 2D layer. Concurrently, the graphene cap is n-doped, while the $Ag_{(2)}$ phase dopes the graphene slightly more due to the higher areal density of silver atoms compared to the $Ag_{(1)}$ phase. Spectroscopic imaging ellipsometry (SIE) supports the semiconducting feature of the Ag-SiC system in both phases while capturing differences in the optical absorption. Second harmonic generation (SHG) measurements demonstrate a three-order magnitude difference in the second-order susceptibility.

**Intercalation of 2D Silver in various types of epitaxial graphene**

Two-dimensional (2D) Ag is synthesized via the CHet method, following the procedure described by Briggs *et al.* Epitaxial graphene (EG) is first prepared from 6H-SiC substrates via high-temperature silicon sublimation, initially forming zero-layer graphene (ZLG), followed by monolayer EG.[39,40] The EG films are then treated with oxygen or helium plasma to introduce defects,[31,41] which facilitates the intercalation of Ag atoms at the EG/SiC interface during thermal evaporation at 900 °C for 1 hour.[12] As illustrated schematically in **Fig. 1a**, all three forms of graphene—ZLG, pristine EG, and plasma-treated EG—enable Ag intercalation, leading to the formation of large-area, crystalline 2D Ag films. These graphene substrates differ in carbon layer structure and thickness as well as by the defect types within the layer. During intercalation, the defects in ZLG and EG tend to self-heal,[31] and ZLG transforms into a free-standing graphene layer after silver intercalation,[12,25] yielding either monolayer or bilayer graphene capping the 2D Ag. The intercalated 2D Ag can form two distinct structural phases, either the $Ag_{(1)}$ phase in $(1 \times 1)$ registry with SiC, and the denser $Ag_{(2)}$ phase.[38]

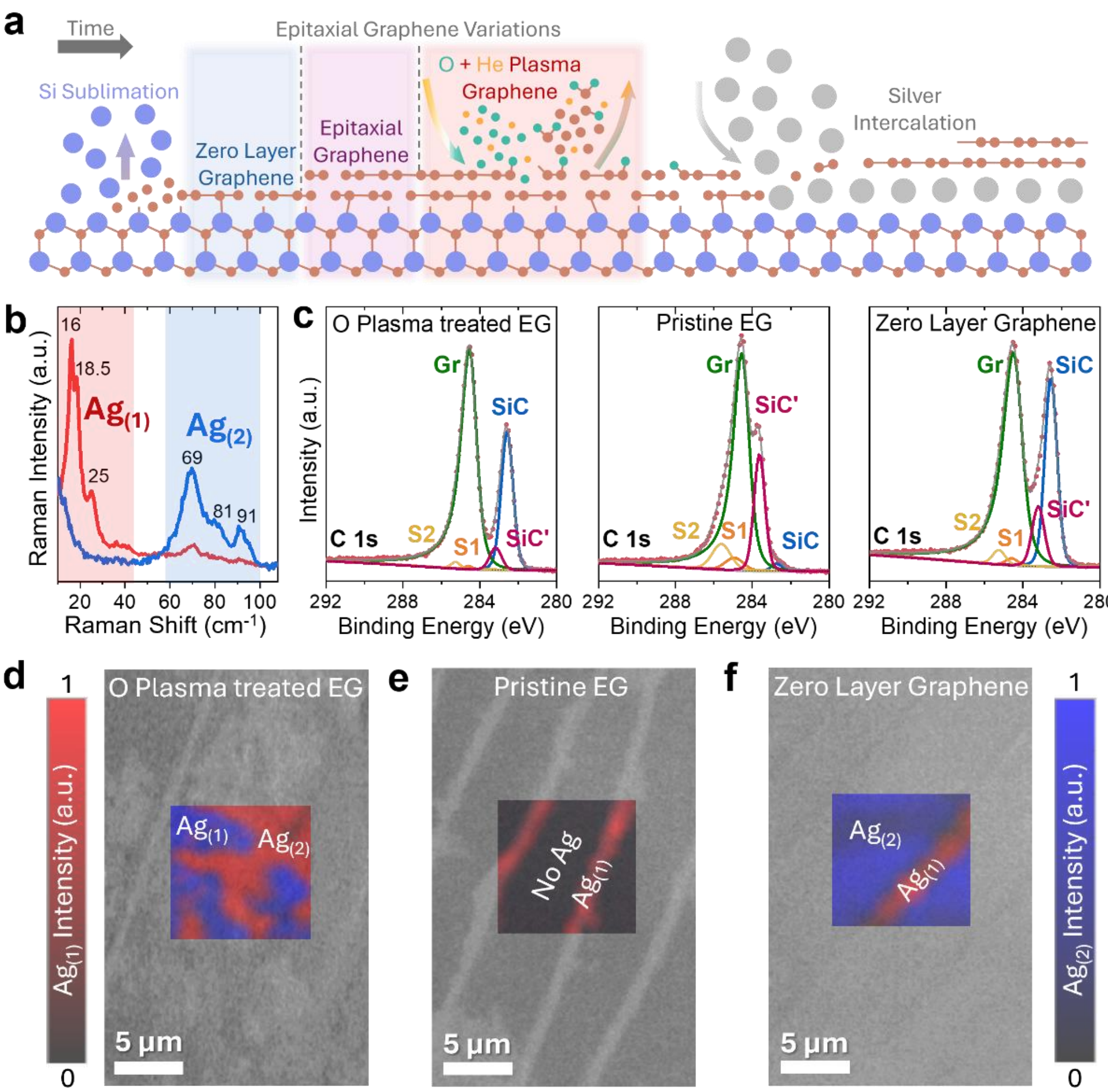

**Fig. 1. Synthesis of the 2D Ag through the CHet method. (a)** Schematic representation of silver intercalation using zero-layer graphene (ZLG), pristine EG, and oxygen plasma-treated EG substrates. **(b)** Characteristic ultra-low frequency (ULF) Raman spectra distinguishing the $Ag_{(1)}$ and $Ag_{(2)}$ phase. **(c)** X-ray photoelectron spectra (XPS) of the C 1s peak of 2D silver samples intercalated on ZLG, pristine EG and O plasma treated EG, indicating varying degrees of buffer layer decoupling as evident by the Gr-SiC peak splitting. **(d-f)** Grayscale optical images of 2D Ag intercalated on **(d)** ZLG, **(e)** pristine EG, and **(f)** O plasma-treated EG, respectively. The insets shows 10×10 $\mu m^2$ Raman map of the ULF region, with mapped Raman intensity corresponding to $Ag_{(1)}$ Raman mode (16, 18.5 and 25 $cm^{-1}$) shown in red, and $Ag_{(2)}$ Raman modes (69, 81, and 91 $cm^{-1}$) shown in blue, corresponding to the spectra in panel (b). Color bars represent the relative Raman intensities for each phase ($Ag_{(1)}$ – red and $Ag_{(2)}$ – blue).

The intercalation of Ag is effectively identified by the emergence of distinct ultra-low frequency (ULF) Raman modes unique to each phase (**Fig. 1b**).[37] The $Ag_{(1)}$ phase exhibits sharp Raman peaks near 16, 18 and 25 $cm^{-1}$, whereas the $Ag_{(2)}$ phase displays three prominent features at 69, 81 and 91 $cm^{-1}$. These ULF Raman modes arise from shear and out-of-plane vibrations involving interactions between the intercalated silver atoms and the underlying silicon bonds, similar to observations in other 2D metals such as Ga and In.[26] Additionally, X-ray photoelectron spectroscopy (XPS) characterization of the carbon 1s spectra (Fig. 1c) confirms the decoupling of the ZLG layer by the intercalant, as indicated by the characteristic splitting between the $sp^2$ carbon (284.5 eV) typical for graphene (Gr), the SiC substrate peaks (282.6 eV) position due to band bending and the reduction in the S1 (284.6 eV) and S2 (285.2 eV) buffer layer peaks.[11] A separation greater than 1.9 eV was observed for intercalated 2D silver on both ZLG and oxygen plasma treated EG, consistent with effective decoupling of the ZLG from the SiC to quasi-freestanding monolayer graphene (QFMLG) and complete Ag intercalation. In contrast, pristine EG samples exhibited a smaller C 1s separation of 1.76 eV, with remnant buffer peaks (S1, S2) and a pronounced SiC' peak from the unintercalated regions of SiC without band bending at 283.6 eV, signifying incomplete intercalation and highlighting the critical role of defects in facilitating Ag incorporation.[31,42,43] The evolution of the C 1s and Si 2p spectra for these samples is shown in **Fig. S1**. The Ag 3d XPS and Ag MNN Auger spectra (**Fig. S2**) further confirm the presence of intercalated 2D Ag in the ZLG and oxygen plasma-treated EG samples.[44]

$Ag_{(1)}$ and $Ag_{(2)}$ exhibit markedly different optical absorption behaviors, as measured via differential reflectance spectroscopy. The $Ag_{(1)}$ phase shows two distinct absorption maxima at approximately 470 nm and 640 nm, whereas the $Ag_{(2)}$ phase features a broad absorption peak centered near 570 nm (**Fig. S3**). These features are associated with localized-plasmon-like interband transitions, characteristic of intercalated 2D metal systems.[45] The resulting contrast in the dielectric functions associated with the absorption behavior enables rapid, phase-sensitive identification under an optical microscope (**Fig. 1d-f**). Using the standard CHet intercalation method on oxygen plasma-treated EG substrates, we achieve complete silver coverage across the 5×10 $mm^2$ samples, with both $Ag_{(1)}$ and $Ag_{(2)}$ phases present. Phase distribution is confirmed by Raman mapping of their respective ULF signatures (**Fig. 1d**), revealing approximately equal proportions of each phase. Correlative Raman and optical imaging shows that $Ag_{(2)}$ regions have a consistently stronger reflective component in the visible range than $Ag_{(1)}$ due to their differing absorption (**Fig. S3**). When the intercalation duration is reduced from 60 to 40 minutes, we observe preferential formation of $Ag_{(1)}$ along terrace edges, with incomplete intercalation and small, mixed-phase domains across the terrace surface (**Fig. S3**). These observations suggest that $Ag_{(1)}$ is the kinetically favored phase, nucleating preferentially at terrace edges in oxygen plasma-treated EG substrates.

Defect density in the graphene overlayer plays a critical role in initiating and guiding silver intercalation at the EG/SiC interface. To elucidate this role, we performed silver intercalation on a pristine EG sample without any plasma treatment, which is expected to exhibit a very low density of defects. Optical imaging and Raman mapping of the ULF region reveal only limited intercalation (< 10%), with minimal to no 2D Ag coverage observed on the terraces (**Fig. 1e)**. Even in the absence of intentional defect introduction, localized intercalation of the $Ag_{(1)}$ phase is observed at terrace edges, where thicker graphene layers typically reside.[40,46] This behavior is consistent with the well-established understanding that epitaxial graphene nucleates at SiC terrace edges, which are inherently rich in structural defects.[46] These results further support the speculation that $Ag_{(1)}$ is the kinetically favored phase that forms first, and that specific defect types, such as those naturally occurring at SiC step edges, play a pivotal role in directing nucleation.

Intrinsic $sp^3$-type defects in ZLG enable preferential formation of the $Ag_{(2)}$ phase during intercalation. While ZLG exhibits a graphene-like structure, approximately one-third of its carbon atoms are covalently bonded to the underlying SiC substrate through $sp^3$ hybridization.[47–49] This bonding configuration introduces a high density of intrinsic $sp^3$-type defects,[50] which are known to facilitate intercalation without the need for ex-situ defect generation.[11] As shown in **Fig. 1e**, optical imaging and ULF Raman mapping reveal extensive intercalation of the $Ag_{(2)}$ phase across ZLG terraces, with localized $Ag_{(1)}$ formation at terrace edges, underneath multilayer graphene. The dominant presence of $Ag_{(2)}$ on ZLG suggests that specific defect types may preferentially promote the nucleation and stabilization of this higher-order phase. Notably, utilizing molecular beam epitaxy (MBE) – an ultra-high vacuum (UHV) technique - in combination with ZLG grown in Argon atmosphere using and RF (radio frequency) furnace leads only to the formation of $Ag_{(1)}$.[25] In contrast, the CHet-based intercalation consistently yields $Ag_{(2)}$ (**Fig. S4**), indicating that differences in intercalation procedures (near atmosphere[31,51] vs MBE[25]), ZLG defect density (ZLG grown with RF or ordered ZLG, **Fig. S4a**) and exposure to the ambient environment impact the defect density and chemistry (e.g. adsorption of molecules), and thus strongly influence phase formation.

**Phase selective growth of 2D Ag through graphene defect engineering**

Tailoring defect types in epitaxial graphene through plasma treatment enables controlled phase selection of 2D Ag. To achieve uniform growth of $Ag_{(1)}$, growth conditions were optimized by systematically modifying the remote plasma treatment of epitaxial graphene (EG) using different gas compositions ($O_2$ and He) and exposure durations. Plasma treatments are known to generate specific defect types in graphene,[41] which can be identified by the D/D′ intensity ratio in Raman spectroscopy.[52] Literature reports associate D/D′ values of ~13, ~7, and ~3 with $sp^3$-type, vacancy-type, and line (boundary) defects, respectively, depending on the plasma chemistry used.[52–56] Raman mapping of treated EG samples revealed distinct trends in D/D′ ratios as a function of plasma parameters (**Fig. 2a**). A 10-second O/He plasma exposure produced a D/D′ ratio of ~9.2, indicative of mixed $sp^3$ and vacancy-type defects. Switching to pure He plasma reduced the ratio to ~7.6, entering the vacancy-type defect regime. Increasing He plasma duration to 2 and 3 minutes further lowered the D/D′ ratios to ~4.6 and ~2.1, respectively, marking a transition toward boundary-defect-type dominance. Corresponding average Raman spectra of plasma-treated EG (**Fig. 2b**), show broadening of the D peak and the emergence of a pronounced D′ shoulder at longer exposures, consistent with increased structural disorder and the formation of extended line defects.[54] In contrast, pristine EG showed minimal D and no visible D′ peak, while zero-layer graphene (ZLG) exhibited a Raman profile characteristic of a well-ordered buffer layer.[49,57,58]

Defect engineering via helium plasma treatment enables controlled, phase-pure growth of 2D $Ag_{(1)}$. Silver intercalation at 900 °C for one hour on plasma-treated EG substrates from **Fig. 2a** resulted in distinct phase outcomes depending on the defect profile. The standard CHet process using O/He plasma treatment[12,31] produced complete intercalation with approximately equal fractions of $Ag_{(1)}$ and $Ag_{(2)}$ phases (**Fig. 2c**). Helium plasma treatment for one-minute increased $Ag_{(1)}$ domain size and coverage to ~60%, suggesting that vacancy-type defects favor $Ag_{(1)}$ nucleation. Extending plasma exposure to two minutes further boosted $Ag_{(1)}$ coverage to ~70%, reinforcing the role of defect type and density. At three minutes of helium plasma exposure, corresponding to the line defect regime, $Ag_{(1)}$ coverage exceeded 95%, yielding large-area, phase-pure 2D $Ag_{(1)}$.

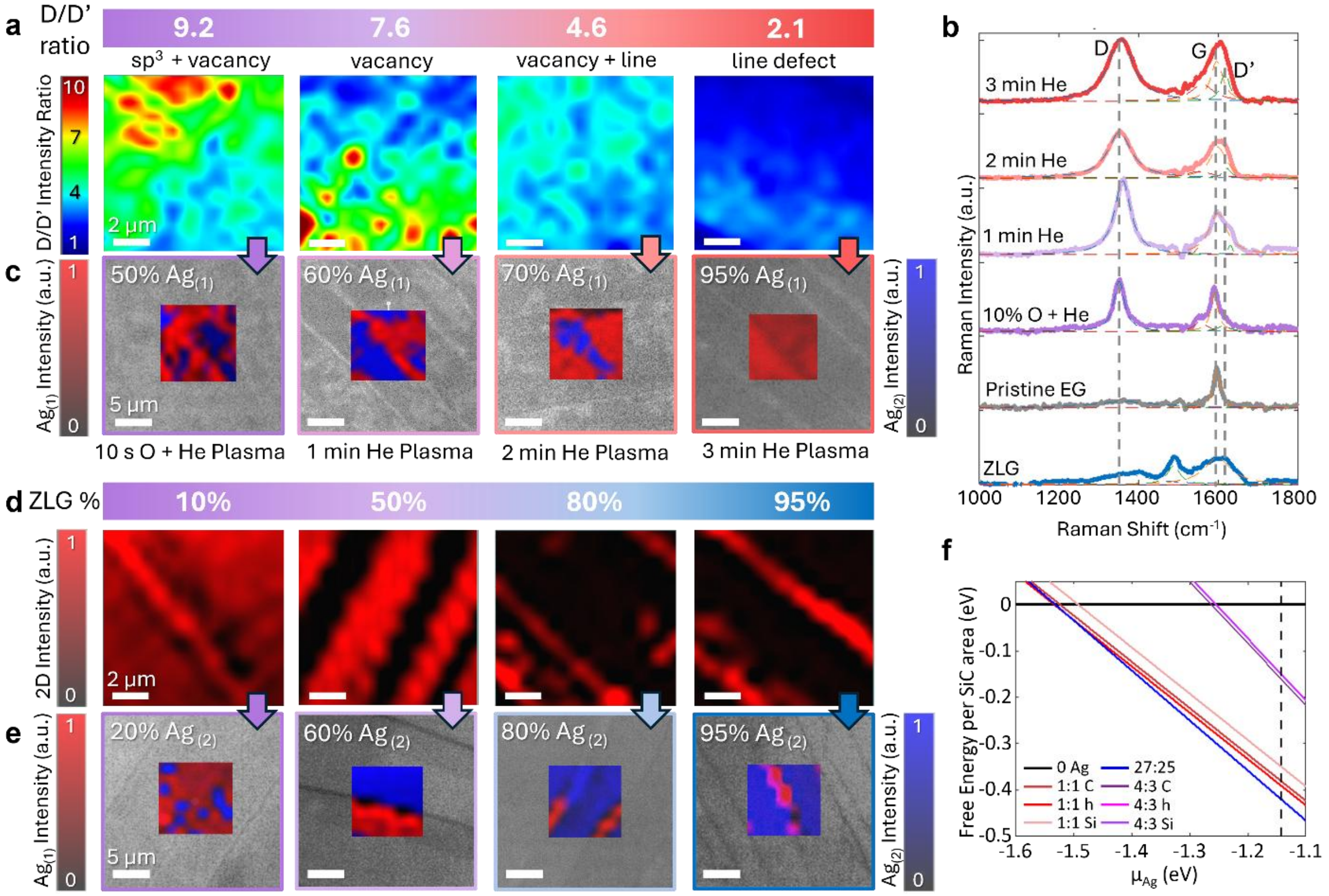

**Fig. 2. Phase engineering of 2D Ag via graphene defect modulation (a)** Raman maps showing the D/D′ intensity ratio in EG substrates treated with 10 s oxygen and helium plasma, and 1, 2, and 3 min helium plasma, highlighting defect type evolution. Scale bars: 2 µm. **(b)** Representative background subtracted Raman spectra of EG substrates (pristine, ZLG, and various plasma treatments), showing fitted D, G, and D′ peaks to assess defect nature and density. **(c)** Grayscale optical images of intercalated samples using EG substrates treated in (a). Insets show ULF Raman maps with $Ag_{(1)}$ and $Ag_{(2)}$ phases represented in red and blue, respectively. Scale bars: 5 µm. **(d)** Raman 2D peak maps of EG for samples with 10%, 50%, 80%, and 95% ZLG coverage, highlighting regions of monolayer EG (red) and exposed ZLG (black). Scale bars: 2 µm. **(e)** Grayscale optical images of Ag-intercalated samples with 20–95% ZLG coverage. Insets show ULF Raman maps identifying $Ag_{(1)}$ and $Ag_{(2)}$ distributions (red and blue, respectively). Scale bars: 5 µm. **(f)** Calculated phase diagram of Ag intercalation (without graphene), showing formation energy per SiC surface unit cello versus Ag chemical potential. "0Ag" refers to the non-intercalated system. The dashed line indicates the chemical potential of bulk Ag. C, Si, and h labels denote Ag atoms on top of carbon, silicon, and hollow sites, respectively (Fig S8). The lowest-energy configuration at each chemical potential identifies the thermodynamically preferred phase, revealing a competition between the 1:1 ($Ag_{(1)}$) and 27:25 ($Ag_{(2)}$) phases. Note: 4:3 Si has a slightly lower free energy compared to 4:3 h site registry.

XPS analysis of this three-minute helium-treated sample before and after intercalation (**Fig. S5**) showed a C 1s peak splitting >1.9 eV and the emergence of $Ag^0$ and Ag-Si 3d peaks, confirming both successful intercalation and graphene-SiC decoupling. These observations suggest that larger, plasma-induced defects—particularly line defects—facilitate direct access for Ag atoms to the underlying SiC, enabling preferential nucleation of the $(1 \times 1)$ $Ag_{(1)}$ phase over the denser $Ag_{(2)}$ structure. This mechanism aligns with earlier observations of $Ag_{(1)}$ formation along terrace edges of pristine EG (**Fig. 1e**), which are naturally enriched with such line defects. Larger defects likely provide pathways for direct Ag–SiC interaction, promoting the formation of the commensurate $Ag_{(1)}$ lattice.[59] Although the as-synthesized $Ag_{(1)}$ film appears uniform and phase-pure, it remains metastable and undergoes a partial transformation (~30% conversion) into the $Ag_{(2)}$ phase over a period of 10 weeks, akin to Ostwald ripening, even when stored in a nitrogen-filled environment at ambient temperature (**Fig. S6**). This long-term phase evolution highlights the thermodynamic preference for $Ag_{(2)}$ and sets the stage for a deeper discussion of the relative stability between the two phases in the following sections.

To promote large-area $Ag_{(2)}$ phase growth, $sp^3$-type defects were targeted as effective intercalation sites by leveraging ZLG substrates. Unlike EG, which requires plasma treatments to introduce $sp^3$ defects, ZLG naturally contains a high density of such defects due to partial covalent bonding between carbon atoms and the underlying SiC substrate.[60] Prior studies have shown that short-duration oxygen plasma treatments can introduce $sp^3$-type defects into graphene,[54] but these often result in insufficient defect densities or transition toward vacancy-type defects at longer exposures. To avoid these limitations, undergrown EG samples with varying degrees of ZLG contribution were synthesized and characterized using 2D Raman peak mapping (**Fig. 2d**). Spatially resolved Raman intensity maps revealed regions with monolayer EG coverage (bright red) and ZLG terraces (dark), corresponding to the presence or absence of the prominent 2D Raman peak. Atomic force microscopy (AFM) adhesion imaging (**Fig. S7**) further confirmed the spatial variation, with ZLG regions exhibiting distinct adhesion contrast.

Modulating the spatial coverage of exposed ZLG and EG enables defect-engineered control over Ag phase intercalation and distribution. Pristine EG substrates with <5% exposed ZLG (>95% EG/ZLG) coverage exhibit minimal Ag intercalation due to a lack of accessible interfacial pathways (**Fig. 1e**). Introducing ~10% ZLG dramatically enhances intercalation and yields predominantly $Ag_{(1)}$, suggesting that $sp^3$-defect-rich ZLG regions trap Ag atoms and facilitate lateral diffusion beneath adjacent EG, enabling widespread metal access. Increasing ZLG coverage to 50% produces coexisting $Ag_{(1)}$ and $Ag_{(2)}$ phases, with $Ag_{(2)}$ localized beneath ZLG and $Ag_{(1)}$ beneath EG, indicating that intercalation pathways extend across phase boundaries while preserving local phase selectivity. Further increasing ZLG coverage to 80% and 95% promotes dominant $Ag_{(2)}$ formation across terraces, with residual $Ag_{(1)}$ appearing along terrace edges and beneath EG regions—echoing trends observed in pristine and plasma-treated EG substrates. These results support the hypothesis that ZLG's small, localized $sp^3$-type defects favor nucleation of the higher-order $Ag_{(2)}$ phase, while larger, extended defects promote $Ag_{(1)}$. Additionally, enhanced charge transfer at the ZLG/SiC interface[61] may further stabilize $Ag_{(2)}$ by modifying interfacial electronic structure. Together, these templating and electronic effects drive rapid growth and stabilization of large-area, phase-pure $Ag_{(2)}$ films on ZLG. However, MBE-based Ag intercalation on RF-grown ZLG still leads to the $Ag_{(1)}$ phase[25] as the RF-grown ZLG is less defective in comparison to ZLG grown by resistive heating.

First-principles calculations evaluating the relative phase stability, finds that $Ag_{(2)}$ is more likely to be thermodynamically favored over $Ag_{(1)}$. The intercalation formation energy density (per SiC surface unit) was computed using:

$$\frac{F}{A} = \frac{E-(E_{0\mathrm{Ag}}+n_{\mathrm{Ag}}\mu_{\mathrm{Ag}})}{n_{\mathrm{SiC}}} \quad (1)$$

where $E$ is the energy of the intercalated system, $E_{0\mathrm{Ag}}$ is the energy of the non-intercalated system, $n_{\mathrm{Ag}}$ is the number of Ag atoms, $\mu_{\mathrm{Ag}}$ is the chemical potential of Ag, and $n_{\mathrm{SiC}}$ is the number of SiC primitive cells on the surface in the computational unit cell. These calculations, performed without the graphene overlayer, are summarized in **Fig. 2f**. For each value of silver chemical potential, the phase with the lowest formation energy density is thermodynamically the most stable. Inside the chemical potential energy window where the intercalation happens, the $Ag_{(2)}$ phase ($n_{\mathrm{Ag}}: n_{\mathrm{SiC}} = 27:25$) is found to be the most stable over a broad range, consistent with the experimentally seen transition from $Ag_{(1)}$ to $Ag_{(2)}$ over time in **Fig. S6**. Additionally, the formation energy density of the $Ag_{(1)}$ phase ($n_{\mathrm{Ag}}: n_{\mathrm{SiC}} = 1:1$) of carbon (C) and hollow-site (h) registries (**Fig. S8**) is very close to that of the $Ag_{(2)}$ phase, which offers the opportunity to tune the relative areal coverage of each phase through graphene defect engineering. In contrast, the 4:3 phase ($n_{\mathrm{Ag}}: n_{\mathrm{SiC}} = 4:3$), which has been proposed in high-throughput computational[62] and experimental[38] studies, exhibits significantly higher formation energy density and is therefore unlikely to form under experimental conditions. While this thermodynamic analysis excludes the effects of graphene or entropy, it offers a semi-quantitative framework that aligns well with the observed phase evolution trends.

The thermodynamic stability of intercalated 2D Ag phases arises from a balance between Ag–Ag cohesion and Ag–SiC substrate interactions, which can be interpreted by comparing Ag–Ag bond lengths across different packing configurations (**Fig. S9**). Bulk FCC Ag exhibits a bond length of 2.88 Å in its (111) plane. In the commensurate (1:1) $Ag_{(1)}$ phase, the Ag–Ag bond length increases to 3.1 Å, matching the Si–Si spacing in the SiC substrate and indicating stronger Ag–SiC bonding that stabilizes this phase despite weaker Ag–Ag interactions. In contrast, the (27:25) $Ag_{(2)}$ phase shows a shorter Ag–Ag bond length of 2.98 Å, closer to the bulk value, suggesting that enhanced Ag–Ag interactions drive the formation of this higher-order phase. The previously proposed 4:3 phase,[38,62] while offering better geometric commensurability than the (27:25) phase, exhibits a significantly short Ag–Ag bond length of 2.68 Å, making it energetically unfavorable.

In addition to these thermodynamic trends, kinetic and structural considerations play a critical role in determining phase formation. As shown in **Fig. S10**, Ag clusters with diameters around 12 Å fail to stabilize the $Ag_{(2)}$ configuration. The high edge-to-area ratio in such small clusters diminishes the stabilizing Ag–Ag interactions needed for $Ag_{(2)}$, making it less favorable during the early nucleation stage. The $Ag_{(1)}$ phase, by contrast, forms more readily due to stronger Ag–SiC interfacial interactions. Thus, under growth conditions involving rapid intercalation—such as regions with line defects (**Fig. 2c**) or terrace edges (**Fig. 1e**)—the system preferentially forms $Ag_{(1)}$. These observations highlight a defect- and kinetics-mediated pathway for phase selection in 2D confined metals.

**Phase-dependent structural and electronic characterization of 2D Ag**

Structural characterization confirms the distinct lattice symmetries and superstructures of the $Ag_{(1)}$ and $Ag_{(2)}$ phases. **Fig. 3a** shows low-energy electron diffraction (LEED) patterns for the two phases, $Ag_{(1)}$ in the upper half, $Ag_{(2)}$ in the lower half, both with the typical first order diffraction, i.e., (10) spots originating from the primitive cells of graphene and SiC (indicated by white and orange circles, respectively). The $Ag_{(2)}$ phase in the lower half exhibits a distinct LEED pattern (**Fig. 3a** (lower half)) with a number of additional spots. Most prominently, six satellite spots arise around the graphene (10) spots (blue circles) corresponding to an approximate (6.25×6.25) grid with respect to graphene. We attribute these to a buckling or a double diffraction superstructure of the graphene on top of the $Ag_{(2)}$ structure in close correspondence with the STM findings (see below). In the $Ag_{(1)}$ pattern (upper half), this spot group is also visible but much weaker, owing to the low percentage residual $Ag_{(2)}$ contribution in this sample. Otherwise, the $Ag_{(1)}$ pattern contains no additional diffraction features, since the $Ag_{(1)}$ phase should be nearly commensurate with the substrate (Ag:SiC≈1:1).[11,25,36] Notably, in the $Ag_{(2)}$ pattern also faint superstructure spots of the $(6\sqrt{3}\times6\sqrt{3})R30°$ supercell of epitaxial orientation of graphene on SiC (green circles) are visible, that arise from the presence of residual non-intercalated areas (see also XPS in **Fig. 1c**). They are absent in the $Ag_{(1)}$ pattern, either due to more complete intercalation or the damping through the bilayer EG cover. Last but not least, we observe five spots in the central region between the electron gun and the graphene (10) spot three of which are part of the $(6\sqrt{3}\times6\sqrt{3})R30°$ superstructure (green circles), and two of which correspond to a (5×5) mesh with respect to SiC.[11,25,36] All these superstructures align with the geometry of the $Ag_{(2)}$ phase (**Table S1**, **Fig. S11**), which consists of a $(3\sqrt{3}\times3\sqrt{3})R30°$ Ag lattice commensurate with a (5×5) SiC unit cell, also referred to as the "27:25" structure due to its Ag:SiC atomic ratio. The 6.25×6.25 graphene supercell originates from commensuration between the Ag and graphene layers, while the $(6\sqrt{3}\times6\sqrt{3})$ SiC superstructure results from the lattice match between graphene and SiC. Schematic atomic structures of both phases are illustrated in **Fig. 3b**: the $Ag_{(1)}$ phase forms a (1×1) triangular lattice aligned with the SiC substrate, analogous to the 1:1 structures observed in CHet-grown Ga and In,[31] while the $Ag_{(2)}$ phase adopts a higher-order $(3\sqrt{3}\times3\sqrt{3})R30°$ configuration over a (5×5) SiC supercell.

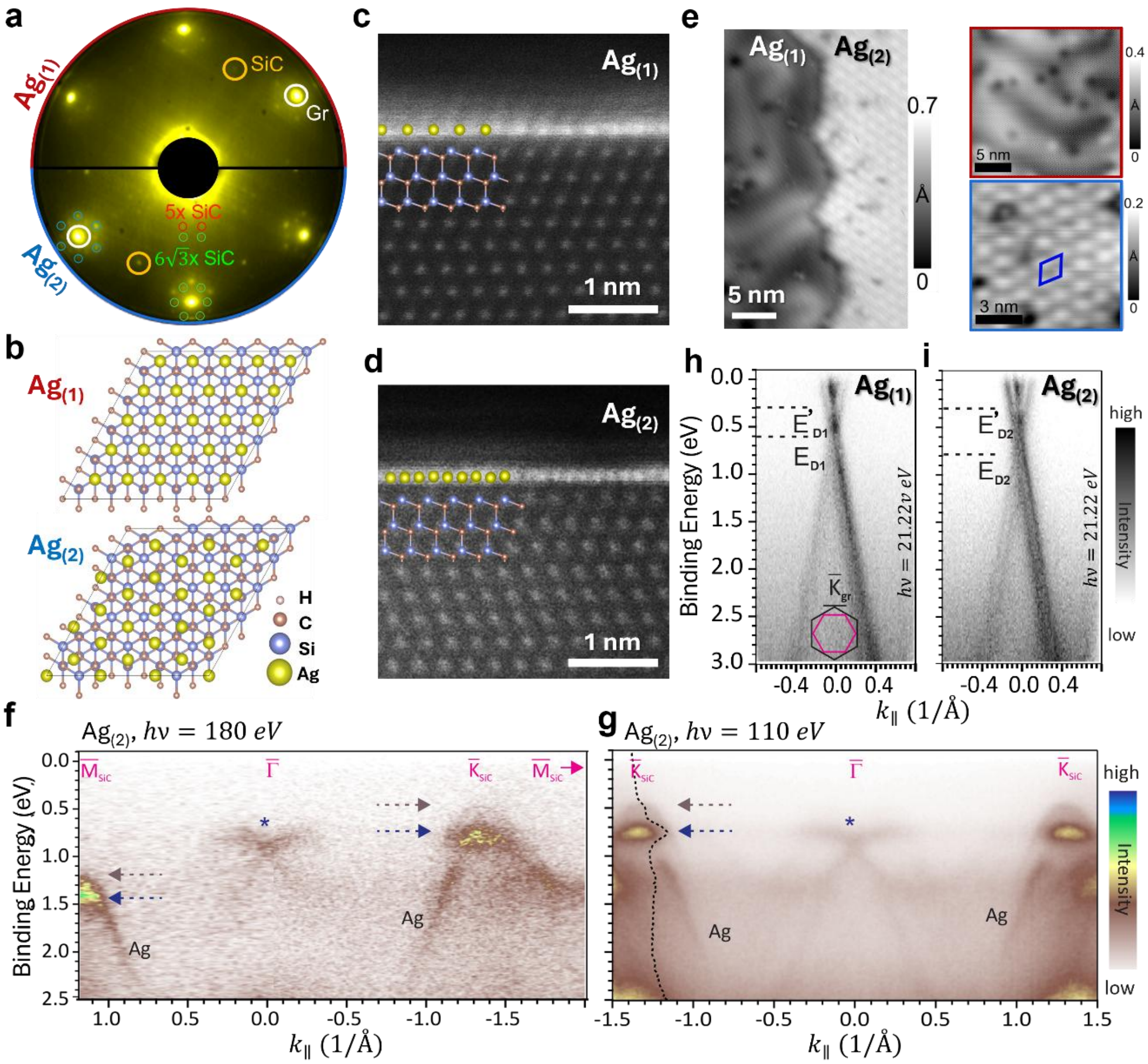

**Fig. 3. Phase-dependent structural and electronic characterization of 2D Ag. (a)** Low-energy electron diffraction (LEED) patterns of the $Ag_{(1)}$ and $Ag_{(2)}$ phase. **(b)** Atomic models of the $Ag_{(1)}$ and $Ag_{(2)}$ phases, illustrating their respective 1:1 and 27:25 commensuration with the SiC substrate (shown without the graphene layer) (c, d) Cross-sectional transmission electron microscopy (TEM) images of the **(c)** $Ag_{(1)}$ and **(d)** $Ag_{(2)}$ phases, with corresponding models overlaid. **(e)** Large-area scanning tunneling microscopy (STM) image acquired at 2 V, showing a boundary between $Ag_{(1)}$ and $Ag_{(2)}$. High-resolution STM images of each phase are outlined in red ($Ag_{(1)}$) and blue ($Ag_{(2)}$). Superstructure shown by the blue rhombus is ~14.7 Å. **(f,g)** High-resolution ARPES maps for $Ag_{(2)}$–QFMLG show $Ag_{(2)}$-band dispersion along **(f)** $\overline{M_{SiC}\Gamma K_{SiC}}$ (180 eV photons) and **(g)** $\overline{K_{SiC}\Gamma K_{SiC}}$ (110 eV photons), with the energy distribution curve at $\overline{K_{Ag}}$ overlaid in **(g)**. Brown and blue arrows in (f, g) mark characteristic bands of $Ag_{(1)}$ and $Ag_{(2)}$, respectively; blue asterisks in (g, h) denote Ag bands backfolded via Umklapp scattering from the graphene BZ. **(h,i)** p-band dispersion in **(h)** $Ag_{(1)}$ quasi-free-standing bilayer graphene (QFBLG) and **(i)** $Ag_{(2)}$ quasi-free-standing monolayer graphene (QFMLG), recorded at $\overline{K_{gr}}$ perpendicular to $\overline{\Gamma K_{gr}}$ direction, measured by ARPES using lab-based He Iα radiation. Dirac points are labeled $E'_{D1(2)}$ and $E_{D1(2)}$ for $Ag_{(1)/((2))}$.

Cross-sectional transmission electron microscopy (TEM) confirms the monolayer nature and atomic registries of both phases. In the $Ag_{(1)}$ phase, the silver atoms exhibit a nearly 1:1 registry with the underlying Si atoms (**Fig. 3c**), while the $Ag_{(2)}$ phase is more densely packed, matching the 27:25 atomic configuration (**Fig. 3d**). The overlayed corresponding structural models align well with the observed atomic positions. Large-area TEM imaging and energy-dispersive X-ray spectroscopy (EDS) maps (**Fig. S12**), further confirm the presence of continuous, monolayer 2D Ag films. Interlayer spacing between graphene and the intercalated silver reveal distinct differences between the two phases: 0.32 nm for $Ag_{(2)}$ and 0.47 nm for $Ag_{(1)}$ (**Fig. S12**). This substantial difference suggests a stronger interaction between graphene and the $Ag_{(2)}$ phase, tentatively due to larger charge transfer from $Ag_{(2)}$ to the graphene (**Fig. 3f, Fig. S13)**, indicating that the graphene overlayer may play an active role in stabilizing specific structural phases during or after intercalation.

Correlative multimodal characterization using Raman, SEM, AFM, and AES confirms structural and electronic distinctions between $Ag_{(1)}$ and $Ag_{(2)}$. These measurements validate differences in packing density, strain/doping in overlying graphene, and silver–graphene interaction strength. A detailed analysis is provided in the Supplementary Information (Note, **Figs. S14–S15**).

Scanning tunneling microscopy (STM) reveals atomic-scale structural and electronic properties of intercalated 2D Ag. Large-area STM topography acquired at +2 V (bias at sample, **Fig. 3e**) clearly resolves adjacent domains of the $Ag_{(1)}$ and $Ag_{(2)}$ phases, separated by a well-defined boundary. Epitaxial graphene is observed atop both phases but becomes visible only at low bias voltages near the Fermi level due to its weak coupling with the underlying Ag (**Fig. S16**). The $Ag_{(1)}$ phase exhibits a non-periodic structure with elongated bright and dark stripes, resembling a herringbone-like pattern in Au(111). We tentatively attribute this modulation to in-plane stress variations within the $Ag_{(1)}$ layer, possibly arising from Ag atoms alternating between hollow and carbon-site positions beneath graphene, driven by small energy differences between these configurations (**Fig. S8**).[63]

In contrast, the $Ag_{(2)}$ phase displays a bright-dot superstructure with a periodicity of ~14.7 Å (highlighted by a blue rhombus), which aligns closely with the 1×1 graphene lattice and six times is commensurate to six times the graphene unit cell (14.76 Å). This periodic modulation is attributed to a moiré pattern resulting from lattice mismatch and angular misalignment between the 27:25 $Ag_{(2)}$ superstructure and the graphene overlayer. The resulting moiré leads to long-range electronic potential modulation at the graphene/Ag interface. Notably, the $Ag_{(2)}$ domains appear slightly elevated relative to $Ag_{(1)}$, though this apparent height difference is bias-dependent. Point defects are also observed in both phases, likely originating from Ag vacancies in the intercalated layer or Si adatoms exchanged with the uppermost SiC substrate. Differential conductance (dI/dV) spectroscopy reveals distinct electronic features of the intercalated 2D Ag phases. Both $Ag_{(1)}$ and $Ag_{(2)}$ exhibit a steep increase in conductance at large positive bias (**Fig. S16**), diverging from the density of states expected for pristine graphene,[64] with Ag(2) showing conduction band onset at ~0.2 eV higher energy than Ag(1). This behavior likely originates from contributions of Ag $d_x$-type orbitals.[62]

In angle-resolved photoemission spectroscopy (ARPES) measurements, the presence of sharp Ag(1) (**Fig. S17a**) and Ag(2) (**Fig. S17b**) bands confirms their well-ordered nature after intercalation. The Ag(1) band (**Fig. S17a**), closely matching previously reported data,[25,36] indicates the semiconducting character. The Ag(2) band structure, recorded along the $\overline{M_{SiC}\Gamma K_{SiC}}$ direction using 180 eV photons, is shown in **Fig. 3f**. The absence of any band crossing the Fermi level further confirms the semiconducting nature of Ag(2). Band splitting in Ag(2) (highlighted by blue and brown arrows) is evident at both the $\overline{M_{SiC}}$ and $\overline{K_{SiC}}$ points. To characterize this splitting at $\overline{K}_{SiC}$, ARPES data were collected along the $\overline{M_{SiC}\Gamma K_{SiC}}$ direction using an excitation energy of 110 eV (**Fig. 3g**), which enhances the Ag photoemission cross-section. The energy distribution curve (EDC) (black dashed line in **Fig. 3g**) at $\overline{K}_{SiC}$ shows two peaks at 0.50 eV (brown arrow) and 0.75 eV (blue arrow) binding energy (BE). The weaker peak at 0.50 eV aligns well with the Ag(1) band maximum,[25,65] while the 0.75 eV peak corresponds to the Ag(2) valence band maximum (VBM). This reveals an energy offset of approximately 250 meV between the Ag(2) and Ag(1) VBMs. The faint bands marked by blue asterisks in Fig. 3(f) and 3(g) near the Γ point correspond to Umklapp-scattered Ag valence bands arising from the graphene Brillouin zone.[24,25]

**Fig. 3h-i** presents ARPES measurements (lab-based He Iα) of the Dirac cones originating from the decoupled graphene layer, recorded at the graphene K point ($\overline{K}_{gr}$) after Ag(1) and Ag(2) intercalation, respectively. Here, $K_{||}$ is oriented perpendicular to the $\overline{\Gamma K_{gr}}$ direction, as indicated in the inset of **Fig. 3h**. In both datasets, two distinct sets of Dirac cones are observed. For the Ag(1)-intercalated sample, (**Fig. 3h**), the Dirac points (DPs) appear at approximately 0.3 eV BE ($E'_{D1}$) and 0.6 eV BE ($E_{D1}$). In the Ag(2) sample (**Fig 3i**), the corresponding Dirac points are labeled as $E'_{D2}$ and $E_{D2}$, respectively. The presence of these dual Dirac cones is more clearly resolved in the synchrotron-based high-resolution data for the Ag(2)-intercalated sample (**Fig. S17d**). The lower-binding-energy Dirac point $E'_{D2}$ (weaker band) corresponds to bilayer graphene (BLG) bands, as confirmed by the observable splitting in the lower part of that cone,[66] a feature not resolved in the lab-based ARPES data due to the energy resolution limit. The higher-binding-energy Dirac point $E_{D2}$ is associated with the quasi-free monolayer graphene (MLG) Dirac cone, while the intensity

elongation between 0.8 eV and 1.2 eV arises from electron–plasmon coupling.[25,65,67,68] In the $Ag_{(1)}$ sample, the BLG Dirac cone ($E'_{D1}$) exhibits stronger intensity. Since this sample was prepared from monolayer graphene on SiC, $Ag_{(1)}$ intercalation likely leads to decoupling of a BLG layer formed during intercalation, whereas the weaker MLG feature ($E_{D1}$) results from decoupling of residual buffer-layer patches. In contrast, the $Ag_{(2)}$ sample—intercalated into a zero-layer graphene substrate—shows a dominant Dirac cone from the quasi-free-standing mono layer graphene (QFMLG) layer after intercalation, while the weaker BLG feature originates from intercalation within overgrown graphene patches.

Notably, the binding energy of $E_{D2}$ is approximately 0.1 eV higher than that of $E_{D1}$, indicating enhanced n-type doping in the ($Ag_{(2)}$–QFMLG) layer. The charge carrier density (n) of the $Ag_{(2)}$–QFMLG layer was estimated using Luttinger's theorem, $n = A/\pi^2$, where A represents the Fermi surface area.[69,70] Neglecting the trigonal warping of the $\pi^*$ band Fermi pocket, it was approximated as a circle with diameter $k_F$, corresponding to the peak separation in the MDC curve at the Fermi level (**Fig. S17e(ii)**) for the $Ag_2$–QFMLG $\pi^*$ bands. With $k_F$=0.274 $Å^{-1}$, this yields n≈ $6.0 \times 10^{13} cm^{-2}$, which is higher than the previously reported value for $Ag_{(1)}$–QFMLG.[36,65] This confirms the stronger n-doping in the $Ag_{(2)}$–QFMLG layer, resulting from increased charge transfer from the $Ag_{(2)}$ layer to the graphene.

**Phase-dependent optical properties of 2D Ag**

The phase-engineered intercalation of 2D Ag enables tunable optoelectronic properties, particularly through phase-dependent optical absorption. To quantify this, we employed spectroscopic imaging ellipsometry (SIE) to extract the complex dielectric functions ($\varepsilon = \varepsilon_1 + i\,\varepsilon_2$) of the $Ag_{(1)}$ and $Ag_{(2)}$. SIE measures changes in the polarization state of monochromatic light reflected from the thin film system, yielding ellipsometric angles Ψ and Δ via the complex reflectance ratio $\rho = r_p/r_s = \tan\Psi * e^{i\Delta}$, where $r_p$ and $r_s$ are the amplitude reflection coefficients of p-, and s-polarized light, respectively.[71] Using a CCD-camera together with suitable lens in the detection arm of the ellipsometer system, hyperspectral maps of Ψ(λ) and Δ(λ) were obtained with sub-micron resolution. **Fig. 4a** shows Ψ maps at 450 nm (~2.76 eV) for $Ag_{(1)}$ and $Ag_{(2)}$ dominant structures highlighting the strong contrast between both phases in the reflection matrix indicating sizeable changes in their respective dielectric response.

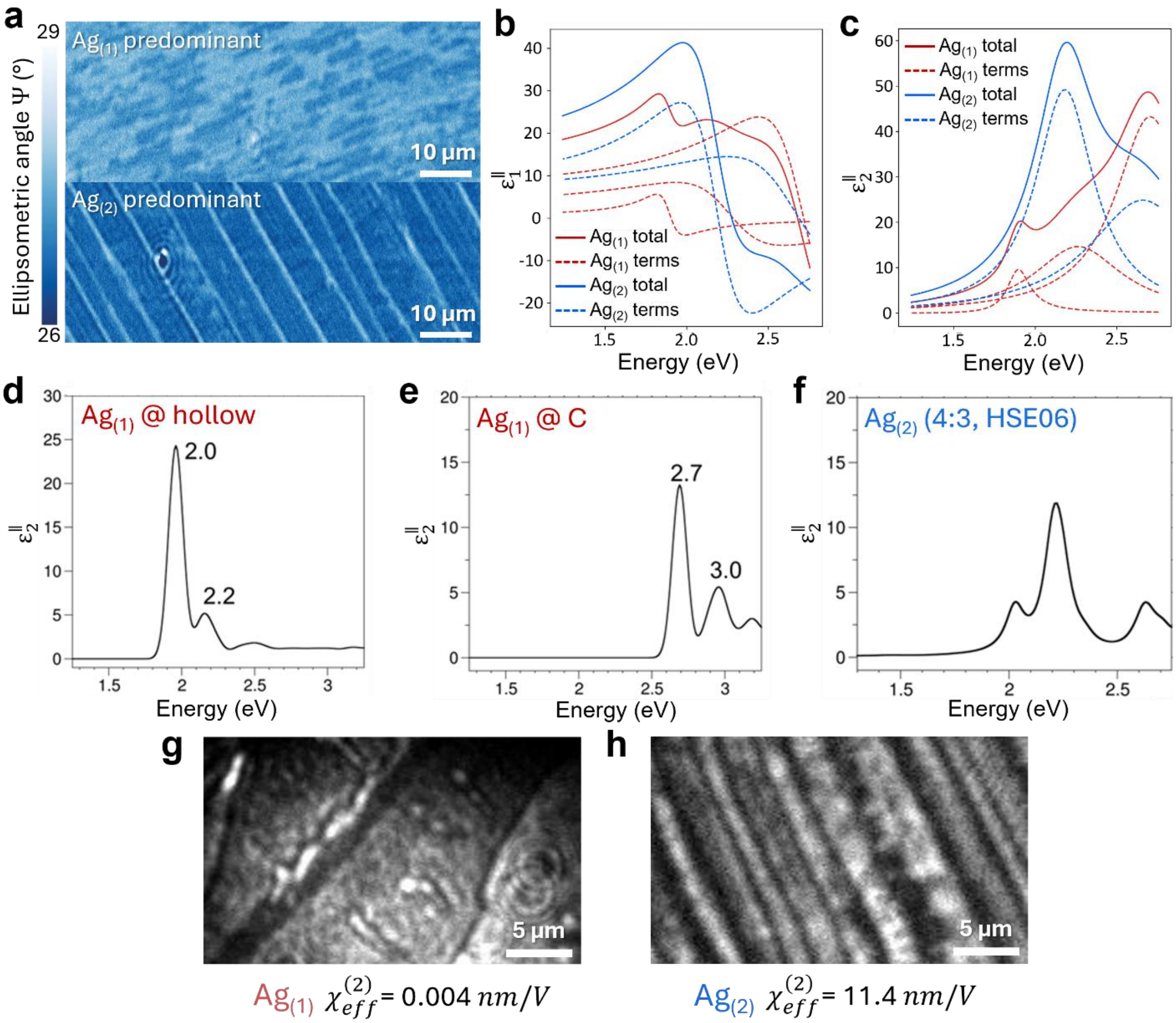

**Fig. 4. Phase-dependent optical properties and characterization of 2D Ag. (a)** Spectroscopic imaging ellipsometry (SIE) Ψ-maps at 450 nm (~2.76 eV) for $Ag_{(1)}$ and $Ag_{(2)}$ dominant samples, showing high ($Ag_{(1)}$) and low ($Ag_{(2)}$) optical contrast regions. **(b)** Real and **(c)** Imaginary part of the dielectric function ($\varepsilon = \varepsilon_1 + i\,\varepsilon_2$) obtained from homogenous regions of $Ag_{(1)}$ predominant sample, shown as red solid lines and from homogenous regions of $Ag_{(2)}$ predominant sample, shown as blue solid lines. Individual terms of the parametrized absorption peaks are shown as dashed lines. **(d-f)** Theoretically calculated optical spectra for **(d)** $Ag_{(1)}$ on hollow site ($Ag_{(1)}$(h)), **(e)** $Ag_{(1)}$ on carbon site ($Ag_{(1)}$(C)) and **(f)** $Ag_{(2)}$. The optical spectra in (d) and (e) include excitonic effects within the GW-BSE approximation, while the optical spectrum in (f) is an estimate obtained using density-functional theory within the HSE06 approximation, applied to the 4:3 model for $Ag_{(2)}$. Second harmonic generation (SHG) images of **(g)** $Ag_{(1)}$ and **(h)** $Ag_{(2)}$ dominant samples with their extracted effective second order susceptibility ($\chi^{(2)}_{eff}$) values, showing strong contrast due to phase-dependent nonlinear optical response.

Ellipsometric spectra ($\Psi(\lambda)$, $\Delta(\lambda)$) extracted from representative regions of the $Ag_{(1)}$ dominant sample were fitted using a multilayer optical model using regression analysis (**Fig. S18**). Very similar results are obtained for the $Ag_{(2)}$ sample following the same approach. The resulting dielectric functions ($\varepsilon$), shown in **Fig. 4b–c**, display clear differences between the real ($\varepsilon_1$) and imaginary part ($\varepsilon_2$) of the dielectric function between $Ag_{(1)}$ and $Ag_{(2)}$. In the loss term ($\varepsilon_2$), $Ag_{(1)}$ exhibits a strong interband resonance at 2.7 eV (~460 nm) and two weaker peaks at 1.9 eV and 2.3 eV, whereas $Ag_{(2)}$ shows a dominant feature at 2.2 eV (~560 nm) and a smaller peak near 2.7 eV.

We perform state-of-the-art calculations for the optical spectra in the $Ag_{(1)}$ system, including the effects of electron-hole interactions for the 2D semiconductor, using the GW-BSE approximation.[72] The calculated $\varepsilon_2$ spectrum is shown in **Fig. 4d** for $Ag_{(1)}$ at the hollow-site registry ($Ag_{(1)}$(h)), and in **Fig. 4e** for $Ag_{(1)}$ at the Carbon-site registry ($Ag_{(1)}$(C)) (see **Fig. S8**). Prominent excitonic peaks are observed at ~2.0 eV and ~2.7 eV for $Ag_{(1)}$(h) and $Ag_{(1)}$(C), respectively. The theoretically-predicted exciton resonances are consistent with the peaks obtained from ellipsometry – a strong interband resonance at 2.7 eV (attributed to $Ag_{(1)}$(C)), and weaker peaks at 1.9 eV and 2.3 eV (attributed to $Ag_{(1)}$h)). This agreement confirms the coexistence of both registries in $Ag_{(1)}$, consistent with their similar formation energies (**Fig. 2f, S8**). We note that the electron-hole interactions dramatically change the low-energy optical spectra (**Fig. S19**), due to the semiconducting and confined nature of the 2D Ag, which has an indirect quasiparticle gap of ~1.0 eV in both registries (**Fig. S20, Table S3**), in stark contrast to other 2D metals grown by CHet, such as 2D Ga,[45] which is metallic. The interband contributions to the lowest-energy exciton resonances are shown in **Fig. S19** – that in $Ag_{(1)}$(h) arises mainly from transitions between the highest occupied and lowest unoccupied bands at the K and M pockets of the Brillouin zone while that in $Ag_{(1)}$(C) arises due to transitions between the highest occupied and lowest unoccupied bands localized mainly close to K. The differences observed in the exciton energies partially reflects the differences in band gaps at the K point – ~2.6 eV for $Ag_{(1)}$(h) and ~3.1 eV for $Ag_{(1)}$(C) structures.

For $Ag_{(2)}$, we provide an estimate for the optical absorption spectra by using as a model the 4:3 Ag:SiC structure (2×2 Ag on $\sqrt{3}\times\sqrt{3}$ SiC)[62] that approximates the electronic environment of (27:25) $Ag_{(2)}$ while maintaining the 30° lattice alignment. We do not take into account excitonic effects in predicting the optical spectra of $Ag_{(2)}$, but use standard density functional theory (DFT) with the hybrid HSE06 exchange-correlation functional, which is commonly used to estimate optical gaps (see also **Fig. S21**).[73,74] Although we do not expect the predicted optical absorption spectrum to quantitatively describe the measured $\varepsilon_2$, the peak positions obtained using this approach can be indicative of the experimental features, and the predicted peak at ~2.2 eV is in reasonable agreement with experiment (**Fig. 4c**).

Beyond linear optics, the broken inversion symmetry at the SiC/Ag interface enables strong second-order nonlinear optical responses unlike bulk Ag.[75] **Fig. 4g** presents second harmonic generation (SHG) microscopy images of $Ag_{(1)}$ and $Ag_{(2)}$ dominant samples, with step and terrace morphology. From the intensity of the images, the extracted sheet susceptibilities ($\chi^{(2)}_{sheet}$) are $0.001 \pm 0.0001$ nm$^2$/V for $Ag_{(1)}$ and $3.2 \pm 0.08$ nm$^2$/V for $Ag_{(2)}$, corresponding to effective susceptibilities ($\chi^{(2)}_{eff}$) of $0.004 \pm 0.0005$ nm/V and $11.4 \pm 0.3$ nm/V, respectively, based on monolayer thickness from TEM (**Fig. 3c,d**) using the sheet model.[76] $Ag_{(2)}$'s $\chi^{2}_{eff}$ is comparable to values for 2D Ga and In,[75] while $Ag_{(1)}$ shows a three-orders-of-magnitude lower response. This stark difference is attributed to stronger Ag–Si interactions in $Ag_{(1)}$, which stiffen the electron cloud and reduce polarizability. These results reveal that precise defect engineering not only governs 2D Ag phase formation but also modulates optical absorption, nonlinear susceptibility, and polarization response—providing a powerful route for designing atomically thin metasurfaces for nanophotonics and ultrathin waveguides.

## Conclusion

This study demonstrates a robust, phase-selective synthesis route for atomically thin silver via defect-modulated confinement heteroepitaxy, enabling deterministic control over the formation of $Ag_{(1)}$ and $Ag_{(2)}$ phases. We uncover a direct correlation between defect type in the graphene overlayer and the resultant ratio between 2D-Ag phases, offering mechanistic insight into the thermodynamic and kinetic pathways governing phase evolution with the possibility to expand this finding to other 2D metal allotropes. The ability to reproducibly engineer structural phase, doping levels, and linear and nonlinear optical responses in 2D-Ag heterostructures unlocks new opportunities in quantum sensing, plasmonics, and tunable photonic systems. Collectively, these findings position intercalated 2D metals as a promising platform for designer quantum materials and next-generation nanophotonic devices.

## Acknowledgements

A.J., M.W-J.L, A.V., N.S., K.K., S.H., V.H.C. and J.A.R. acknowledge the support from National Science Foundation (NSF) Award No. DMR-2011839 (through the Penn State MRSEC–Center for Nanoscale Science). B.Z., N.S. and V.H.C. thank the support from Two-Dimensional Crystal Consortium-Materials Innovation Platform (2DCC-MIP) under NSF cooperative agreement no. DMR-2039351. S.D. and U.St. acknowledge support by the Deutsche Forschungsgemeinschaft (DFG, German Research Foundation) within the Research Unit FOR5242 through Project Sta 315/13-1. MAX IV Laboratory is acknowledged for time on beamline BLOCH under proposal 20241262. We are also thankful to the beamline staff at BLOCH for their support. Research conducted at MAX IV, a Swedish national user facility, is supported by the Swedish Research council under contract 2018-07152, the Swedish Governmental Agency for Innovation Systems under contract 2018-04969, and Formas under contract 2019-02496. K.A.U. and S.Y.Q. acknowledge support from the Ministry of Education in Singapore under grant no. MOE-T2EP50223-0010. The computational work for this article was partially performed on resources of the National Supercomputing Centre, Singapore (https://www.nscc.sg). S.H. acknowledges the support from National Science Foundation (grants ECCS-2246564 and ECCS-1943895) and the Welch Foundation (award C-2144).

## Methods

### *ZLG and EG synthesis*

Zero-layer graphene (ZLG) and epitaxial graphene (EG) were synthesized on the Si-terminated (0001) face of semi-insulating 6H-SiC substrates (Coherent Corp.) via thermal silicon sublimation. The substrates were first annealed at 1400 °C for 30 min in a 10% $H_2$/Ar gas mixture at 700 Torr to remove polishing damage and surface contaminants. ZLG formation was then achieved by annealing the substrates at 1600 °C in 700 Torr of pure Ar for 30 min. Partial ZLG coverage was obtained by tuning the annealing temperature and duration. Monolayer EG was synthesized by further annealing at 1800 °C in 700 Torr of pure Ar for 20 min.

### *ZLG growth by RF heating in Ar atmosphere*

In this method, ZLG samples were prepared by heating an on-axis, Si-terminated, H-etched 6H-SiC(0001) (SiCrystal GmbH) substrate in an argon (impurity level < 10 ppm) atmosphere[40] in near-atmospheric Ar pressure (approx. 600 Torr), with an argon flow rate in the range of 0.2 slm. During annealing, the temperature was maintained at 1444 °C for 4 minutes and 30 seconds. This yields several-µm wide terraces, uniformly covered with the $(6\sqrt{3}\times6\sqrt{3})R30°$ superstructure characteristic for perfect epitaxy of graphene on SiC.

*Plasma Treatment and 2D Silver intercalation*

To facilitate silver intercalation, epitaxial graphene (EG) surfaces were first activated using remote helium plasma treatment in a Pie Scientific Tergeo plasma cleaner. The treatment was carried out under 15 sccm He flow at 6 Torr pressure, 15 W RF power, and a 0.1 duty cycle for 1, 2, or 3-minute. In the case of oxygen plasma treated graphene, monolayer EG samples were subjected to a reactive oxygen plasma etch (150 sccm $O_2$, 50 sccm He, 500 mTorr, 50 W, 1 minute) using a Tepla M4L plasma chamber.

Silver intercalation was performed in a one-inch outer diameter quartz tube furnace (Thermo Scientific Lindberg/Blue M Mini-Mite). ZLG or plasma-treated EG/SiC substrates ($1 \times 0.5$ cm$^2$) were placed face-down inside an alumina crucible (Robocasting Enterprises) positioned above 80 mg of high-purity silver powder (99.99%, Sigma-Aldrich). The tube was evacuated to ~5 mTorr, leak-checked, and then filled with Ar to 500 Torr. The system was ramped to 900 °C at 50 °C/min under continuous Ar flow (50 sccm), held at temperature for 1 h (unless otherwise noted), and then cooled to room temperature under inert atmosphere.

*Raman Spectroscopy and Microscopy*

Raman spectroscopy was performed using a Horiba LabRAM system equipped with a 532 nm excitation laser and an ultra-low frequency (ULF) notch filter to suppress the Rayleigh line. For ULF Raman measurements, spectra were acquired using a 300 grooves mm$^{-1}$ grating with a total integration time of 10 s, averaged over two accumulations and a laser power of ~4 mW. Raman mapping was conducted using the SWIFT ultra-fast imaging mode with a typical spatial resolution of $0.8 \times 0.8$ µm$^2$ per pixel. A 100× objective (NA = 0.9) was used for all measurements, and spectral calibration was performed using a standard Si reference. All spectra were baseline corrected, and for the ZLG/EG Raman spectrum, an additional SiC background subtraction was applied.

Strain and carrier concentration in graphene were extracted via vector decomposition of the Raman G and 2D peak positions. Following background subtraction and peak fitting, spatial maps of the G and 2D peak frequencies were used to decouple contributions from strain and doping based on their known sensitivities to uniaxial strain and charge carrier density. The analysis employed reported Grüneisen parameters and empirical doping coefficients derived from literature.[49,77,78] Data points were plotted in the 2D vs G peak position space, and projections along the iso-strain and iso-doping vectors enabled extraction of relative changes in strain (ε) and electron concentration (n).

*X-Ray Photoelectron Spectroscopy*

X-ray photoelectron spectroscopy (XPS) measurements were performed using a Physical Electronics VersaProbe III system equipped with a monochromatic Al Kα X-ray source (hν = 1486.6 eV) incident at 45° relative to the sample plane. Spectra were acquired using a concentric hemispherical analyzer over a $200 \times 200\ \mu m^2$ analysis area. High-resolution scans of the C 1s, Si 2p, and Ag 3d regions were collected at a pass energy of 29 eV, while the O 1s and Ag MNN Auger regions were recorded at a pass energy of 55 eV. Samples were mounted directly onto a screw-fastened platen without additional adhesives. Spectra were charge-referenced by setting the $sp^2$-hybridized carbon (C 1s) peak to 284.5 eV.

*Atomic Force Microscopy (AFM)*

Atomic force microscopy (AFM) measurements were carried out using a Bruker Dimension Icon system operated in PeakForce Tapping mode. A ScanAsyst-Air probe was used with a peak force setpoint of ~700 pN. Topography and adhesion maps were acquired at a scan rate of 1.0 Hz.

*Low Energy Electron Diffraction (LEED)*

Low-energy electron diffraction (LEED) measurements were performed using a Er-LEED system (Specs GmbH). Samples previously exposed to ambient conditions were degassed in UHV (base pressure $< 5 \times 10^{-10}$ mbar) at 130 °C for 25–30 minutes to remove surface adsorbates. LEED patterns were acquired at an incident electron energy of 70 eV.

Scanning Transmission Electron Microscope (STEM)

Cross-sectional lamellae were prepared using a focused ion beam (FIB) lift-out technique in a dual-beam system (Helios Nanolab 660, Thermo Fisher Scientific). Rough trenching to isolate the region of interest was performed at 30 kV with a beam current of 9 nA. Subsequent thinning was carried out at 30 kV and 0.79 nA, followed by final polishing at 5 kV and 0.36 nA to minimize surface damage and amorphization.

High-angle annular dark-field scanning transmission electron microscopy (HAADF-STEM) imaging and energy-dispersive X-ray spectroscopy (EDS) elemental mapping were conducted using a double aberration-corrected FEI Titan3 G2 microscope operated at 300 kV. The HAADF detector had a collection angle of 52–253 mrad. EDS maps were acquired using a SuperX detector in STEM mode.

Image analysis was performed using ImageJ and Gatan DigitalMicrograph. Average intensity line profiles were extracted from selected regions of interest (ROI) across the cross-sections to quantify the interlayer spacing between the Ag and graphene layers. These profiles were averaged across rectangular windows spanning ~10 nm in the lateral direction to minimize noise and highlight periodic contrast features. Relative intensities in the HAADF images were used as a proxy for atomic number (Z) contrast, allowing estimation of the number of Ag atomic layers based on comparative brightness.

*Scanning tunneling microscopy (STM) measurements*

Intercalated Ag was investigated using a cryogenic STM system (Createc) operating at 5 K in UHV. The samples were transported in air and degassed at 150 °C for 30 min in UHV before transferring into the cold STM head held at 5 K. The topography and differential conductance (dI/dV) spectroscopy were obtained using electrochemically etched W tips. Before measurements, the tips were cleaned by $Ne^+$ sputtering and electron beam heating. The tip calibration was carefully done using an Ag (111) crystal until good topography images were achieved and differential tunneling spectroscopy (dI/dV) showed a surface state onset at about −63 mV, typical for Ag. dI/dV spectra were recorded using a lock-in technique with a peak-to-peak bias modulation of 5 mV and a modulation frequency of 675 Hz.

*Secondary Electron Microscopy (SEM)*

Surface morphology was examined using an Apreo 5 scanning electron microscope (Thermo Fisher Scientific) operated in immersion mode. Secondary electron (SE) imaging was performed using the Through-Lens Detector (TLD). Imaging was conducted at magnifications of 2,000× to 10,000×, with an accelerating voltage of 2.00 kV, beam currents ranging from 0.4 to 3.2 nA, and working distances between 2 and 4 mm.

*Auger Electron Spectroscopy (AES) mapping*

AES measurements were performed using a Physical Electronics VersaProbe III system equipped with a $LaB_6$ scanning electron microscopy column (1–10 keV) and a concentric hemispherical analyzer. A monochromatic Al Kα X-ray source (hν = 1486.6 eV) was used for XPS-based height alignment, ensuring optimal positioning relative to the analyzer prior to AES analysis. The SEM column and analyzer were oriented at 45° takeoff angles with respect to the sample plane and rotated 90° relative to one another. Auger spectra were acquired using a 10 keV, 2 nA electron beam with an estimated probe diameter of ~200 nm. The analyzer was operated in Fixed Retard Ratio (FRR) Transmission mode.

Spot and line scans were collected in scanned mode. Post-acquisition data processing involved signal differentiation and smoothing using a 5-point Savitzky–Golay filter in *CasaXPS*.[79] Elemental quantification was guided by relative sensitivity factors extracted from XPS spectra collected over the same ~20 μm region. Auger mapping was conducted in unscanned mode using Physical Electronics' three-point intensity calibration. This qualitative method involves subtraction of a linear background across two energy intervals and intensity averaging over a third peak region to yield time-normalized signals. For energy calibration, absolute reference values were taken from Seah et al.[80]

*Angle-resolved Photoelectron Emission Spectroscopy (ARPES)*

ARPES measurements were performed using both lab-based and synchrotron-based systems to probe the graphene π-bands and the electronic band structure of 2D Ag. Prior to the measurement, samples previously exposed to ambient air were degassed at 130 °C for 25–30 minutes in UHV (base pressure $< 5 \times 10^{-10}$ mbar) to remove surface contaminants. Lab-based ARPES measurements were performed at room temperature using a NanoESCA system (Scienta Omicron GmbH) operating in momentum microscopy mode with an extractor voltage of 12 kV. During measurement, single-shot, constant-energy cuts in momentum space (k-space) were obtained from ~ 20 × 15 μm² area of the sample surface, controlled by an adjustable iris aperture. Non-monochromatized, unpolarized He Iα radiation (hν = 21.22 eV) served as the excitation source. The spectrometer was operated with a k-space field of view of 4.3 $Å^{-1}$, enabling full access to the photoemission horizon up to the Fermi level for this photon energy.

High-resolution ARPES measurements were performed at the Bloch beamline of the MAX IV synchrotron (Lund, Sweden) using a DA30-L hemispherical analyzer (Scienta Omicron GmbH). The sample temperature during these measurements was maintained at approximately 20 K. For excitation, photon energies of 40 eV, 110 eV, and 180 eV with linear horizontal polarization were employed. The total energy resolution, including contributions from both the analyzer and the beamline optics, ranged between ~17 meV and ~60 meV depending on the photon energy and experimental configuration. Data were analyzed using Igor Pro 6.37 software.

*Spectroscopic Imaging Ellipsometry (SIE)*

SIE measurements are performed in ambient conditions with a customized EP4 ellipsometer (Park Systems GmbH). Tunable, monochromatic illumination is achieved by a supercontinuum white light laser in combination with acousto-optic tunable filters, yielding a spectral linewidth on the order of 5 nm. The signal is imaged onto a CCD camera using a 50x objective with NA = 0.45, which results in a lateral resolution of ~ 1 μm.

Data is obtained at a fixed angle of incidence of 50° in rotating-compensator ellipsometry (RCE) mode, and with a spectral range of 450 - 1000 nm. Data points are spaced 5 nm apart for a total of 111 points. A beam cutter is included in the excitation path to suppress backside reflections from the transparent substrate. The measured spectroscopic maps are manually corrected for small sample movement during the measurement due to e.g. vibrations. Then, suitable homogeneous regions of interest (ROIs) are chosen to average over, resulting in ellipsometric spectra $\Psi(\lambda)$, $\Delta(\lambda)$. To determine the complex dielectric function for $Ag_{(1)}$ and $Ag_{(2)}$, an optical multilayer model is constructed and fit to the spectra via regression analysis (see supplement for a detailed description).

*SHG*

Polarization-resolved SHG microscopy was implemented following the method from Steves *et al.*[75] 800-nm pulses with 30-nm bandwidth were focused onto the sample with a 0.4 NA aspheric lens from a Ti:sapphire laser (Vitara, Coherent). Polarization of the excitation was adjusted with a halfwave plate placed before the sample, while the emission from the sample was collected with a 1.25 NA oil immersion objective (Nikon). The SHG signal was isolated from the fundamental and multiphoton photoluminescence using a combination of 400-nm bandpass and 650-nm short-pass filters. An analyzer was placed before the detector to determine the polarization of the emission. An electron-multiplying CCD and coupled spectrometer (iXion Ultra 897/Shamrock 303i, Andor) were used to record the SHG image and spectra.

Absolute sheet susceptibilities ($\chi^{(2)}_{sheet}$) were extracted from calibrated CCD photon counts, and effective susceptibilities ($\chi^{(2)}_{eff}$) were calculated using monolayer thickness from TEM in **Fig. 4c-d**, yielding $\chi^{(2)}_{sheet} = 0.001 \pm 0.0001$ nm²/V and $\chi^{(2)}_{eff} = 0.004 \pm 0.0005$ nm/V for $Ag_{(1)}$, and $\chi^{(2)}_{sheet} = 3.2 \pm 0.08$ nm²/V and $\chi^{(2)}_{eff} = 11.4 \pm 0.3$ nm/V for $Ag_{(2)}$ as per Steves *et al.*[75]

*Thermodynamic Calculations*

The relative phase stability, and the Ag cluster calculations are done by density functional theory (DFT) as implemented in VASP[81–84] at the PBE-level[85] with 450 eV cutoff energy, electronic self-consistent convergence at $1.4\times10^{-8}$ eV/atom, and residual force convergence at 0.01 eV/Å on atoms with D3 van der Waals correction.[86,87] The k-point samplings are Γ-centered 13×13×1 for 1×1 SiC unit cells, 8×8×1 for 3×3 SiC unit cells, 3×3×1 for 5×5 SiC unit cells, and Γ-only for 10×10 SiC unit cells. The cell dimension in the z-direction is 20Å. Three layers of SiC are included in the phase stability calculations, while two layers of SiC are included in the cluster calculation. All SiC systems are terminated with H atoms at the opposite site of the Ag layer. The structure visualization is made with VESTA.[88]

*Calculation for Optical Properties*

The optical absorption spectra for the $Ag_{(1)}$ phase were obtained with GW-BSE calculations using the BerkeleyGW code,[72] on top of DFT calculations performed using Quantum ESPRESSO.[89] The DFT starting points utilized the local density approximation for the exchange correlation functional. Optimized norm-conserving Vanderbilt (ONCV) pseudopotentials[90,91] were used, with a plane wave basis set cutoff of 90 Ry. An energy threshold of $10^{-10}$ Ry was used to converge the electronic cycles. The $Ag_{(1)}$ phase was modeled with a six-layered slab of SiC with one layer of Ag atoms stacked over the Si face at the hollow- and C-site registries, while the dangling bonds on the bottom C-face were passivated with hydrogen atoms. A vacuum spacing of ~18 Å was used to reduce interactions between the periodic slabs. A uniform 18×18×1 $k$ -grid was found to be sufficient to converge the total energy. A slab Coulomb truncation approach[92] was used in GW calculations. For calculating the dielectric matrix, we use a cutoff of 40 Ry and a summation over ~1400 conduction bands. The quasiparticle energies were obtained using 1400 conduction bands for the summation of the self-energy for 15 valence and 6 conduction bands over an 18×18×1 $k$-point mesh. The electron-hole interaction kernel was calculated using an 18×18×1 $k$-point mesh, which was then interpolated onto a fine 54×54×1 $k$ -grid for 6 valence and 4 conduction bands, that were used to obtain a converged excitonic absorption spectra.

For the $Ag_{(2)}$ phase, we obtained the optical absorption spectra using the independent particle approximation within DFT. The results in **Fig. 4** are obtained using the 4:3 Ag:SiC structure and the HSE06 exchange-correlation functional. The DFT transition dipole matrix elements and energy eigenvalues were calculated on a 36×36×1 $k$-point mesh with PBE. HSE06 calculations were performed on a coarser 6×6×1 $k$-point mesh, and a 3×3×1 q-point mesh for calculating the Hartree-Fock exchange. Incorporating these corrections in the eigenvalues and rescaling of the transition dipole matrix elements, we obtained the HSE06 optical spectra for the 4:3 Ag:SiC structure, using the 6×6×1 $k$-point mesh. We investigated the effect of the choice of structural models by comparing the optical absorption spectra the 4:3 Ag:SiC structure and the 27:25 Ag:SiC, using the PBE exchange-correlation functional (**Fig. S21**). For the 27:25 Ag:SiC structure, the optical spectra was calculated using a 6×6×1 $k$-point mesh.

# Supporting Information

# Defect-Mediated Phase Engineering of 2D Ag at the Graphene/SiC Interface

Arpit Jain[1], Boyang Zheng[2,3], Sawani Datta[4], Kanchan Ulman[5], Jakob Henz[6], Matthew Wei-Jun Liu[1], Van Dong Pham[7], Wen He[5,8], Chengye Dong[1,3,9], Li-Syuan Lu[1], Alexander Vera[1], Nader Sawtarie[10], Wesley Auker[11], Ke Wang[11], Bob Hengstebeck[11], Zachary W. Henshaw[12], Shreya Mathela[13], Maxwell Wetherington[11], William H. Blades[12], Kenneth Knappenberger[13], Ursula Wurstbauer[6], Su Ying Quek[8,14,15], Ulrich Starke[4], Shengxi Huang[16], Vincent H. Crespi[2], and Joshua A. Robinson*[,1,8,17]

[1]Department of Materials Science and Engineering, The Pennsylvania State University, University Park, PA 16802, USA

[2]Department of Physics, The Pennsylvania State University, University Park, PA 16802, USA

[3]2-Dimensional Crystal Consortium, The Pennsylvania State University, University Park, PA 16802, USA

[4]Max-Planck-Institut für Festkörperforschung, Heisenbergstraße 1, 70569 Stuttgart, Germany

[5]Department of Physics, National University of Singapore, Singapore 117551

[6]Institute of Physics and Center for Soft Nanoscience (SoN), University of Münster, 48149 Münster, Germany

[7]Paul-Drude-Institut für Festkörperelektronik, Hausvogteiplatz 5-7, Leibniz-Institut im Forschungsverbund Berlin e. V., 10117 Berlin, Germany

[8]Department of Materials Science and Engineering, National University of Singapore, Singapore 117575, Singapore

[9]Center for 2-Dimensional and Layered Materials, The Pennsylvania State University, University Park, PA 16802, USA

[10]Department of Chemical and Petroleum Engineering, University of Pittsburgh, Pittsburgh, Pennsylvania 15260, USA

[11]Materials Research Institute, The Pennsylvania State University, University Park, PA 16802, USA

[12]Department of Physics and Engineering Physiscs, Juniata College, Huntingdon 16652, Pennsylvania, USA

[13]Department of Chemistry, The Pennsylvania State University, University Park, PA 16802, USA

[14]Centre for Advanced 2D Materials, National University of Singapore, Singapore 117542, Singapore

[15]NUS Graduate School, Integrative Sciences and Engineering Programme, National University of Singapore, Singapore 117456, Singapore

[16]Department of Electrical and Computer Engineering and the Rice Advanced Materials Institute, Rice University, Houston, TX, USA

[17]Center for Atomically Thin Multifunctional Coatings, The Pennsylvania State University, University Park, PA 16802, USA

[*] Corresponding author: jar403@psu.edu

**Additional X-ray Photoelectron Spectroscopy (XPS) Characterization for Gr/Ag/SiC**

The structural evolution of the 2D Ag samples grown through the confinement heteroepitaxy (CHet) method can be monitored through changes in the high-resolution C 1s and Si 2p core levels measured by x-ray photoelectron spectroscopy (XPS) (**Fig S1**). Prior to intercalation, the zero-layer graphene (ZLG) sample displays a characteristic C 1s spectrum (**Fig S1a**) with SiC substrate peaks at 283.6 eV and buffer-layer S1 and S2 features at 284.6 eV and 285.2 eV, respectively. Additional $sp^2$ carbon and oxygen-functionalized C–O (286.8 eV) and C=O (288.3 eV) peaks reflect the presence of overgrown terrace-edge EG and partial oxidation of the buffer. Pristine epitaxial graphene (EG) samples show (**Fig S1b**) strong $sp^2$ graphene peaks alongside residual buffer components and the SiC substrate signal, while oxygen plasma treatment introduces prominent C–O and C=O features (**Fig S1c**), indicating surface functionalization.[1]

Following Ag intercalation, the C 1s spectra exhibit clear signatures of successful buffer layer decoupling and Ag incorporation. Both ZLG and oxygen-plasma-treated EG samples show significant reduction in the S1 and S2 buffer peaks, complete disappearance of the oxygenated carbon signals due to defect healing,[1] and increased intensity of the $sp^2$ peak. Notably, the SiC substrate peak shifts by ~1 eV to lower binding energy (282.6 eV), consistent with interfacial band bending.[1,2] These changes confirm the formation of an additional quasi-freestanding monolayer graphene (QFMLG) and successful intercalation of Ag at the graphene/SiC interface, further corroborated by Raman mapping (**Fig. 1d–f**). In contrast, pristine EG samples exhibit incomplete intercalation, with persistent S1/S2 buffer peaks and both band-bended (SiC peak at 282.6 eV)) and bulk SiC' (283.6 eV, unintercalated regions of SiC without band bending) peaks, suggesting coexistence of intercalated and non-intercalated regions.[1,3]

The Si 2p core-level evolution complements C 1s results and highlights the interfacial effects of Ag intercalation. All three starting samples display substrate (101.4 eV), surface (101.9 eV), and minor $SiO_x$ (102.6 eV) peaks (**Fig. S1d–i**). After intercalation, ZLG and plasma-treated samples reveal a distinct Ag–SiC signal shifted to ~100.3 eV, attributed to downward band bending from interfacial Ag. A separate SiC' component at 101.0 eV corresponds to regions without Ag, while the $SiO_x$ peak persists due to substrate defects. For pristine EG, the band-shifted Ag–SiC peak appears significantly weaker relative to the unshifted SiC' signal, reinforcing that intercalation is incomplete in this case. Together, the C 1s and Si 2p analyses establish a strong spectroscopic framework for evaluating intercalation quality and structural transformation in 2D Ag heterostructures.

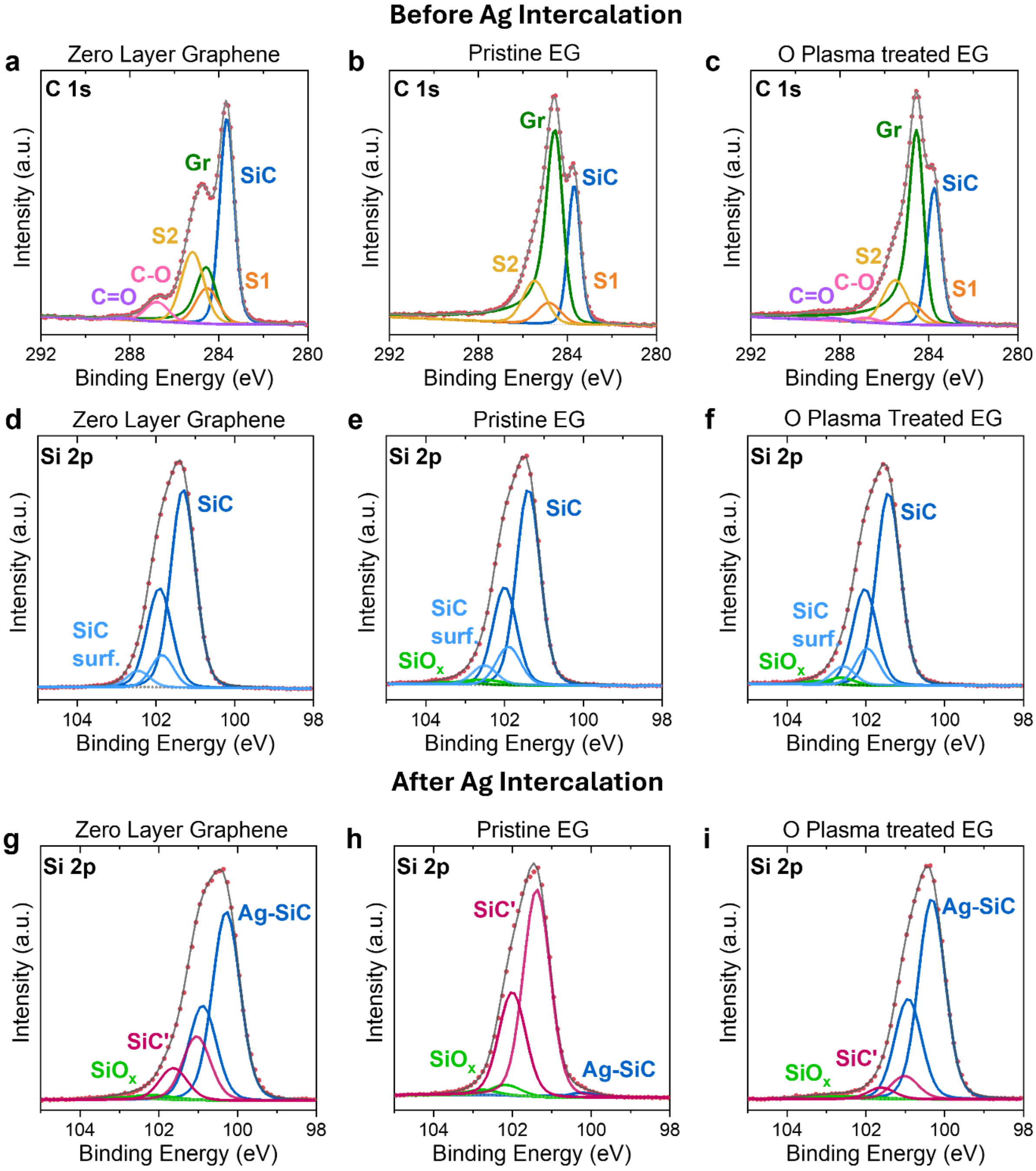

**Fig. S1. Core level XPS analysis for CHet evolution of EG/Ag/SiC.** High-resolution **(a-c)** C 1s and **(d-f)** Si 2p spectra for ZLG, pristine EG, and O plasma-treated EG before Ag intercalation. Corresponding Si 2p spectra after intercalation are shown in **(g–i)** for the same three samples. The spectra reveal interfacial band bending, buffer layer decoupling, and intercalation quality through characteristic shifts and intensity changes in SiC, buffer, and oxide-related peaks.

**Silver XPS and Auger Characterization of Intercalated 2D Ag**

We used Ag 3d core-level X-ray photoelectron spectroscopy (XPS) and Ag MNN Auger spectroscopy to confirm the presence of intercalated 2D silver in Gr/Ag/SiC heterostructures. As a reference, we first measured a sputter-cleaned bulk silver film (**Figs. S2a–b**), which exhibited an Ag $3d_{5/2}$ peak at 368.3 eV, a spin–orbit doublet separation of 6 eV, and a plasmon loss feature at 372.2 eV—values consistent with bulk metallic Ag.[4] The Ag MNN Auger peak appeared at 1128.8 eV binding energy (corresponding to 357.8 eV kinetic energy using a 1486.6 eV excitation source).

To evaluate the bonding character of intercalated silver, we calculated the Auger parameter, defined as the sum of the Ag $3d_{5/2}$ binding energy and the Ag MNN kinetic energy. Pristine silver yields a parameter >725 eV, whereas compound or oxidized Ag results in a lower value.[4] After silver intercalation, all three sample types—ZLG, pristine EG, and oxygen plasma-treated EG—show Ag $3d_{5/2}$ peaks centered at 368.5 eV and a distinct shoulder at ~368.9 eV and an absence of the bulk silver plasmon loss feature (**Figs. S2c–e**). We assign the main $Ag^0$ peak to in-plane Ag–Ag bonding, and the higher binding energy shoulder to out-of-plane Ag–Si covalent interactions. Corresponding Ag MNN peaks (**Figs. S2f–h**) in all samples yield Auger parameters exceeding 725.5 eV.

Among the three intercalated systems, the pristine EG sample exhibited the lowest XPS signal intensity and signal-to-noise ratio, indicating a reduced intercalation efficiency compared to ZLG and oxygen-functionalized EG. This trend further corroborates the correlation between graphene defect density and intercalation quality, as detailed in the main text.

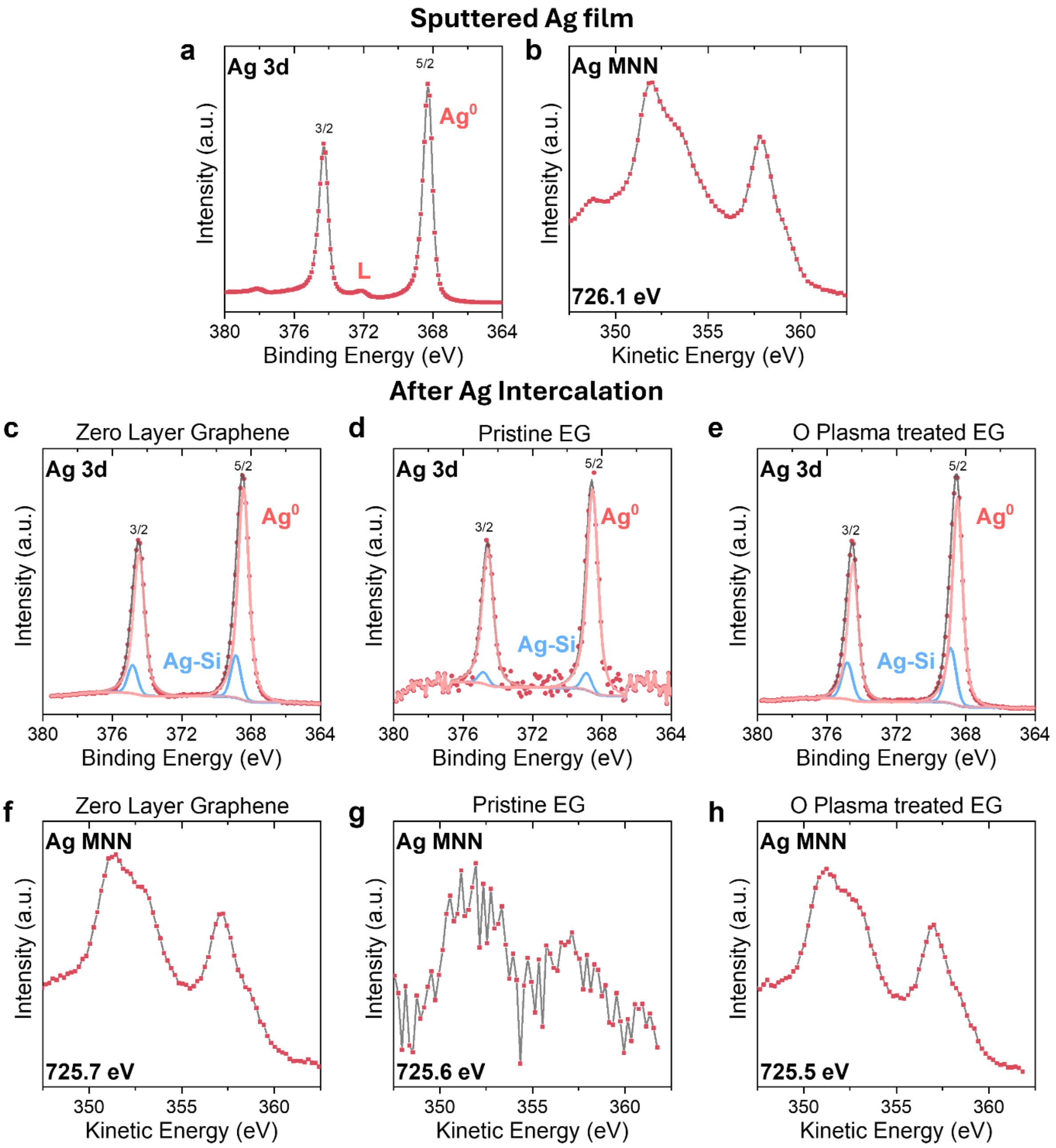

**Fig. S2. Ag 3d XPS and Ag MNN Auger spectra for intercalated Ag in Gr/Ag/SiC system. (a)** High-resolution Ag 3d XPS and **(b)** Ag MNN Auger spectra for a reference Ag foil after $Ar^+$ ion sputtering. High-resolution **(a-c)** Ag 3d XPS and (**f-h)** MNN Auger spectra for intercalated silver in **(c, f)** ZLG, **(d, g)** pristine EG, and **(e, h)** oxygen plasma-treated EG samples, respectively.

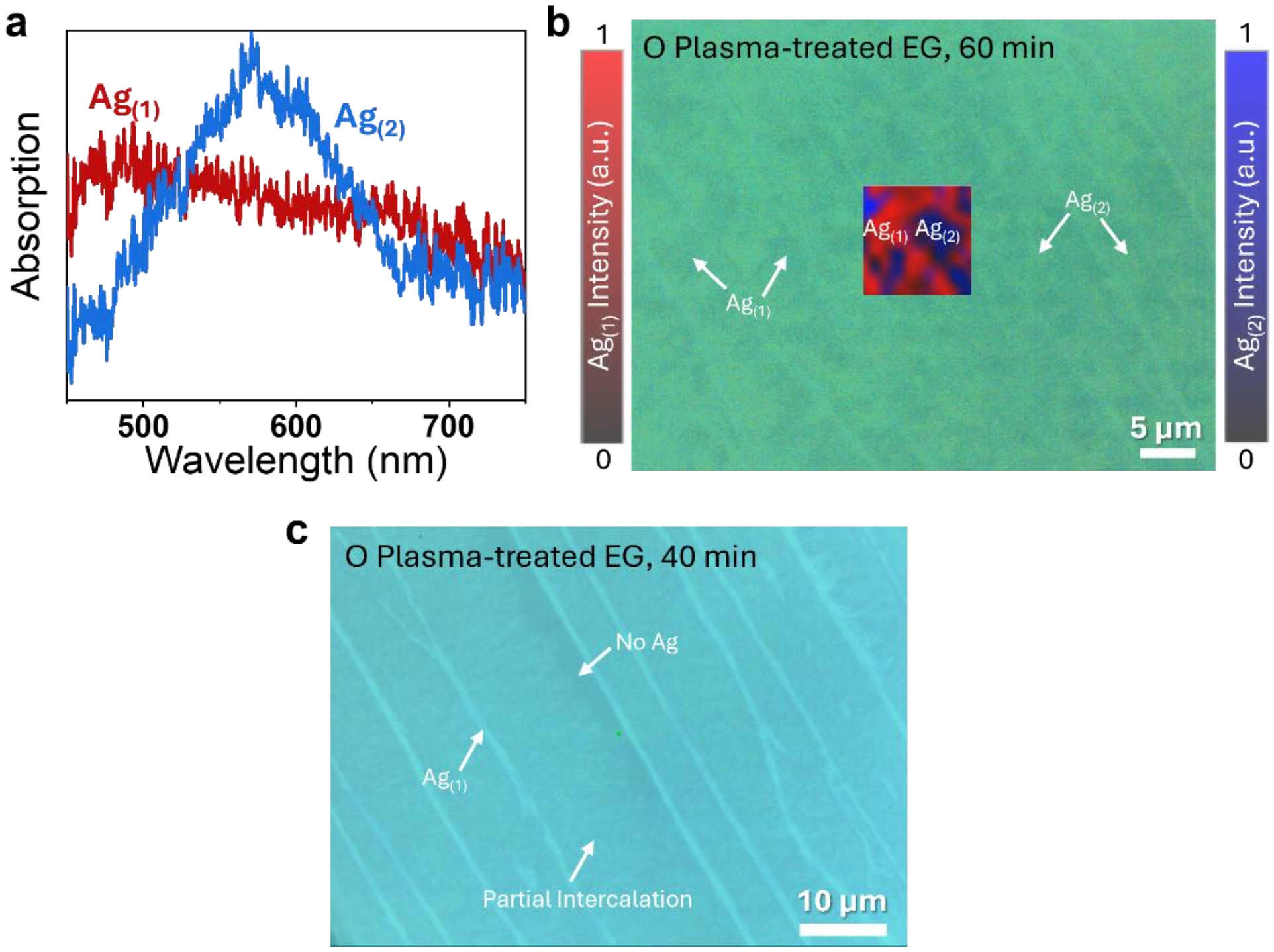

**Fig. S3. Phase-dependent optical absorption and contrast in 2D Ag. (a)** White-light differential reflection spectra ($\frac{\Delta R}{R}(\omega) = \frac{R(\omega)-R_0(\omega)}{R_0(\omega)}$), for $Ag_{(1)}$ and $Ag_{(2)}$, directly reflecting optical absorption. Both spectra show interband transitions in good agreement with Zhang et al.,[9] with experimental and theoretical details described previously in their work.[9] **(b)** Optical micrograph of fully intercalated 2D Ag sample (60-minute intercalation) on oxygen plasma-treated EG. Inset shows 10×10 µm$^2$ Raman map of the ULF region, with mapped Raman intensity corresponding to $Ag_{(1)}$ Raman intensity map highlighting the $Ag_{(1)}$ modes (16, 18.5, 25 cm$^{-1}$, shown in red) and $Ag_{(2)}$ modes (69, 81, 91 cm$^{-1}$, shown in blue). Color bars indicate the relative intensity of each phase's Raman features. (c) Optical micrograph of a partially intercalated sample (40-minute intercalation) on oxygen plasma-treated EG, with regions labeled for $Ag_{(1)}$, partially intercalated, and unintercalated areas.

**Silver intercalation on Zero Layer Graphene prepared in Ultra High Vacuum (UHV)**

ZLG samples grown in Ar atmosphere in the RF-furnace at MPI-Germany, provide a contrasting platform to ZLG synthesized at PSU. Raman spectra (**Fig. S4a**) of the RF-furnace ZLG, after subtracting the SiC background, reveal characteristic signatures of both ordered buffer (sharp 1500 $cm^{-1}$ peak) and disordered buffer regions (broad bands at ~1350 $cm^{-1}$ and 2500–3000 $cm^{-1}$) as reported by Turker *et al.*,[5] indicating lateral phase inhomogeneity within the laser spot size (~1 µm).

Raman mapping analysis confirms the presence of intercalated silver beneath the RF furnace-grown ZLG. As shown in **Fig. S4b**, CHet-grown Ag beneath the RF-furnace grown ZLG sample from MPI exhibits coverage exclusively by the $Ag_{(2)}$ phase, in contrast to pure $Ag_{(1)}$ growth reported for MBE-based intercalation in UHV conditions.[6] Dark contrast regions in the optical image align with undergrown SiC terraces, where neither graphene growth nor Ag intercalation occurs.

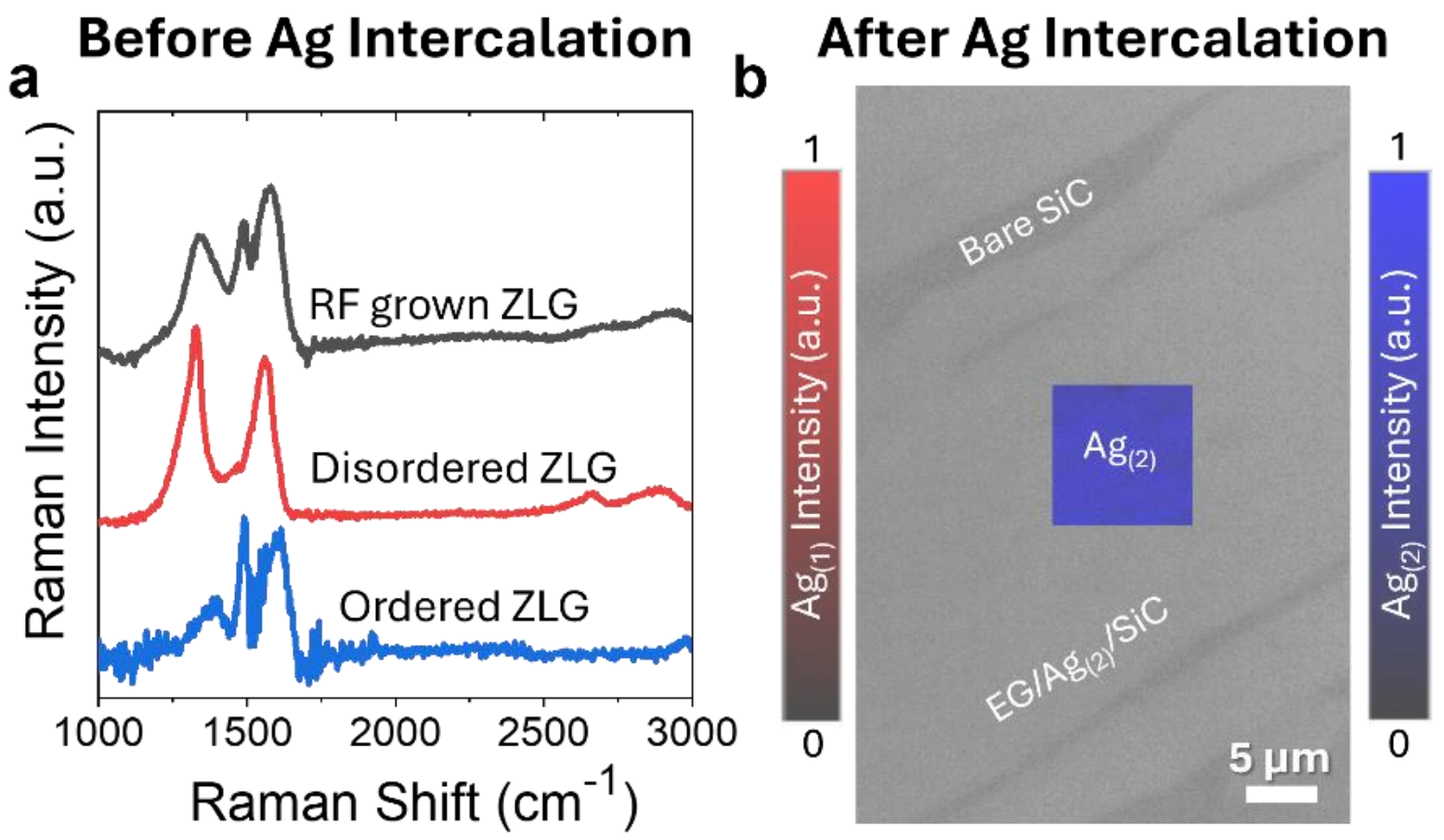

**Fig. S4. Ag CHet-intercalation on RF-furnace grown ZLG. (a)** Representative background-subtracted Raman spectra for RF furnace-grown ZLG from MPI and ordered/disordered ZLG samples from PSU. **(b)** Optical micrograph of Ag CHet-intercalated sample (60 min) on RF grown ZLG, with inset showing a 10 × 10 $\mu m^2$ Raman map of the ULF region. Raman intensities of $Ag_{(1)}$ modes (16, 18.5, and 25 $cm^{-1}$) are shown in red, and $Ag_{(2)}$ modes (69, 81, and 91 $cm^{-1}$) in blue. Color bars represent relative intensities of each phase.

**XPS Characterization of Silver intercalated on Helium Plasma Treated Graphene**

This note discusses the evolution of silver intercalation in helium (He) plasma treated epitaxial graphene (EG) using high-resolution core level X-ray photoelectron spectroscopy (XPS). The intercalation process is tracked through changes in the C 1s, Si 2p, and Ag 3d spectra. Before intercalation, the C 1s spectrum (**Fig. S5a**) reveals peaks for the SiC substrate (283.8 eV), the $sp^2$ graphene signal (284.5 eV), buffer layer components S1 (284.8 eV) and S2, and oxygen-functionalized carbon species, including C–O (285.8 eV) and C=O (288.3 eV). The corresponding Si 2p spectrum (**Fig. S5b**) displays the SiC bulk (101.4 eV) and surface (101.9 eV) peaks, along with a $SiO_x$ shoulder near 102.5 eV.

After Ag intercalation, the C 1s spectrum (**Fig. S5d**) exhibits a clear shift in the SiC peak to lower binding energy (282.5 eV), consistent with downward band bending. Simultaneously, the buffer layer peaks (S1, S2) are significantly suppressed, and a strong $sp^2$ graphene peak re-emerges at 284.5 eV—indicative of buffer decoupling.[1,2] The disappearance of the oxygen-functionalized carbon peaks signals effective graphene healing during intercalation **(ref)**. Si 2p spectra (**Fig. S5e**) further confirm band bending, as the Ag–SiC peak shifts to ~100.2 eV, and a SiC′ peak at 100.9 eV reflects partially unintercalated regions. The Ag 3d core level (**Fig. S5c**) presents peaks at 368.4 eV (Ag-Ag binding, $Ag^0$) and 368.8 eV (Ag–Si bonding), confirming presence of intercalated silver. The Ag MNN Auger spectrum (**Fig. S5f**) yields an Auger parameter exceeding 725 eV, which also verifies the presence of silver in the sample.

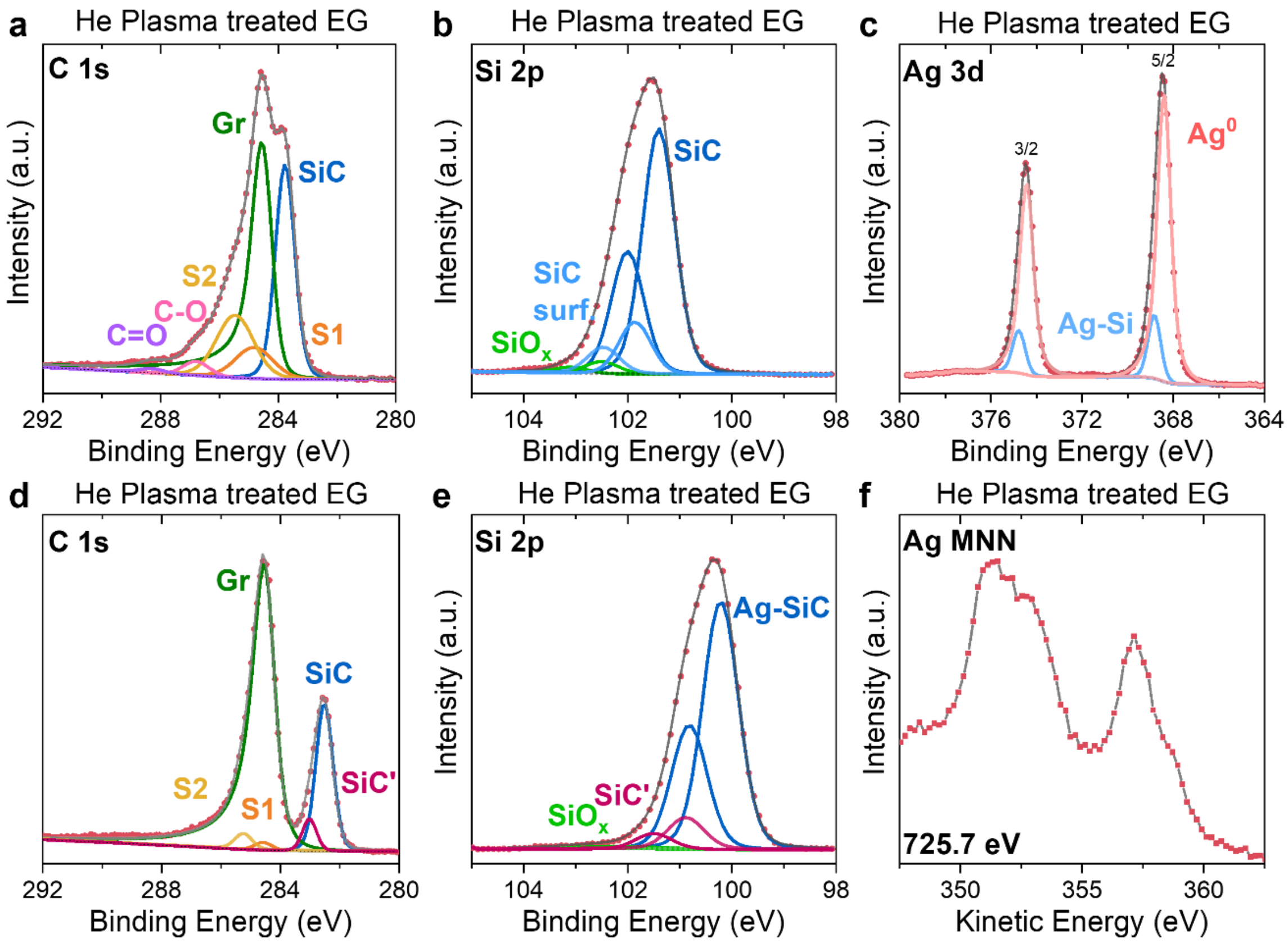

**Fig. S5. XPS characterization of Ag intercalation on helium plasma treated EG.** High-resolution **(a, d)** C 1s and **(b, e)** Si 2p XPS spectrum for 3-minute helium plasma treated EG, before and after Ag intercalation, respectively. **(c)** Ag 3d XPS and **(f)** MNN Auger spectrum for the same sample after intercalation, confirming the presence of Ag and successful interface bonding with SiC.

## Metastability of the $Ag_{(1)}$ phase

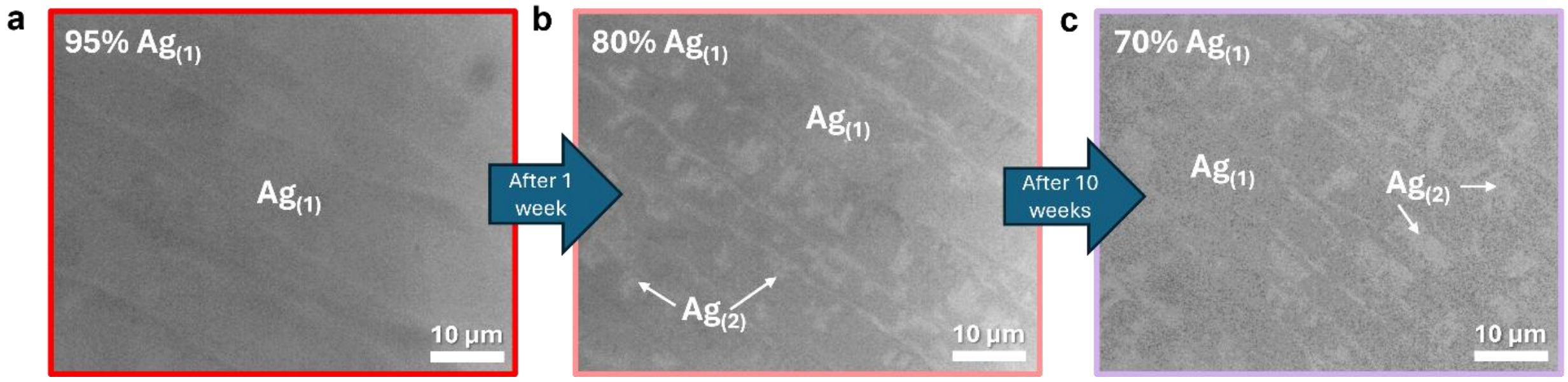

**Fig. S6. Time-dependent conversion of $Ag_{(1)}$ to $Ag_{(2)}$.** Greyscale optical micrographs of a 2D silver sample grown on 3-minute Helium plasma treated EG **(a)** immediately after growth, **(b)** after 1 week, and **(c)** after 10 weeks of storage. Over time, dark-contrast $Ag_{(1)}$ regions progressively convert into bright-contrast $Ag_{(2)}$ domains, with a rapid initial transformation that slows and saturates at ~30% conversion. This behavior experimentally confirms the higher thermodynamic stability of $Ag_{(2)}$ compared to $Ag_{(1)}$. Samples were stored in a nitrogen-filled drawer at room temperature and briefly exposed to ambient conditions during each characterization.

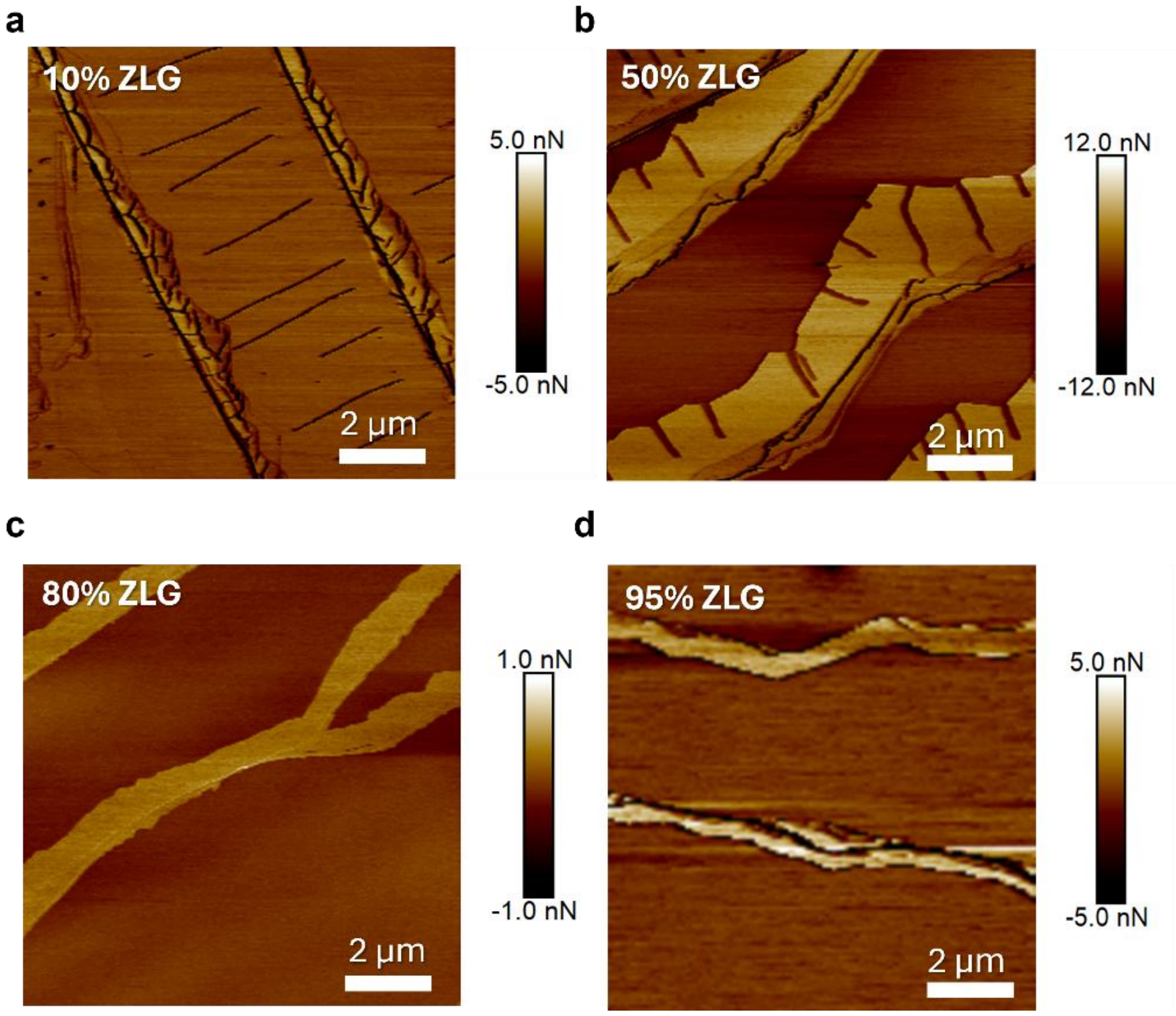

**Fig. S7. Atomic Force Microscopy (AFM) characterization of ZLG coverage.** AFM Adhesion maps for **(a)** 10%, **(b)** 50%, **(c)** 80%, and **(d)** 95% exposed zero layer graphene (ZLG) illustrate a reliable method for quantifying ZLG coverage. Regions with lower adhesion (darker contrast) correspond to ZLG, while higher adhesion (brighter contrast) indicates overgrown epitaxial graphene (EG). These adhesion contrasts provide a facile and spatially resolved approach to estimate ZLG percentage across the surface.

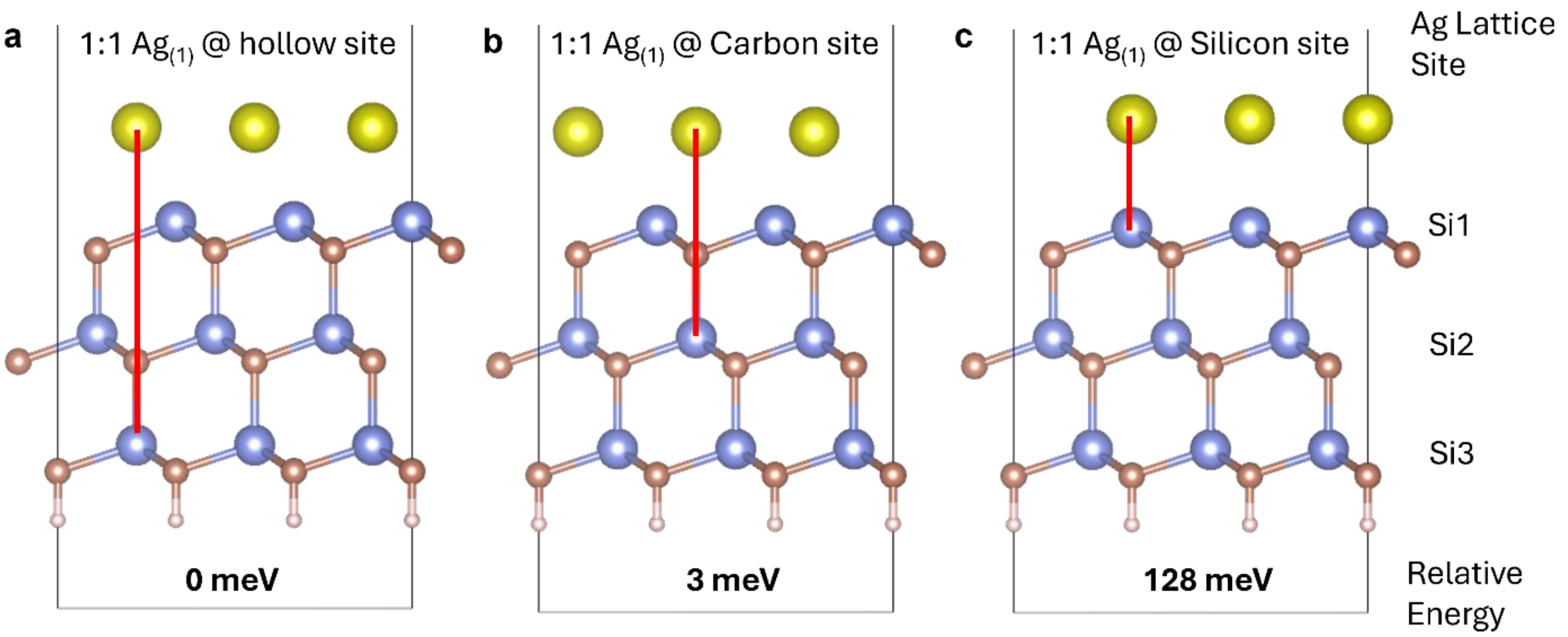

**Fig. S8. Stable lattice registries for $Ag_{(1)}$ over SiC**. Side-view structural models of the three stable registries of the $Ag_{(1)}$ phase over the SiC substrate: $Ag_{(1)}$(hollow), $Ag_{(1)}$(carbon), and $Ag_{(1)}$(silicon). Relative total energies are reported with respect to the hollow-site configuration, which is the lowest-energy and therefore most thermodynamically favorable registry. These configurations illustrate the registry-dependent bonding behavior that contributes to the coexistence and optoelectronic properties of the $Ag_{(1)}$ phase.

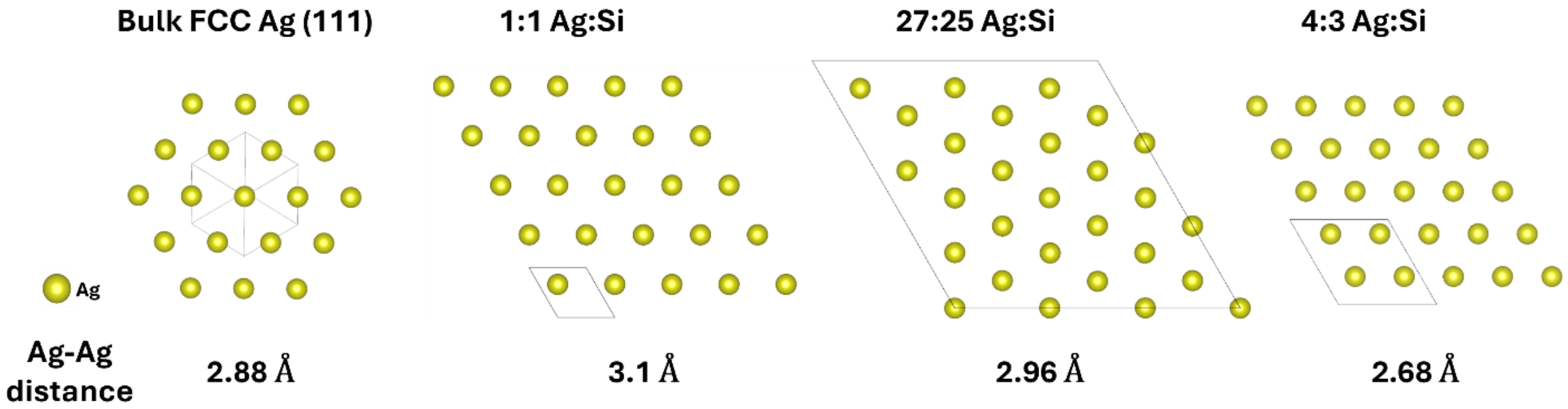

**Fig. S9. Comparison of Ag-Ag interatomic spacing in various packing densities of Ag**. Top-view structural models showing the atomic arrangement of monolayer Ag in three different commensurate phases with the SiC substrate— 1:1 ($Ag_{(1)}$), 27:25 ($Ag_{(2)}$), and (c) 4:3 Ag:Si configurations. These are compared with the bulk Ag FCC (111) plane. The Ag–Ag interatomic distances vary between these configurations, illustrating how 2D confinement and registry with the underlying substrate modulate the atomic packing density of the intercalated silver. Respective unit cells are shown in light grey.

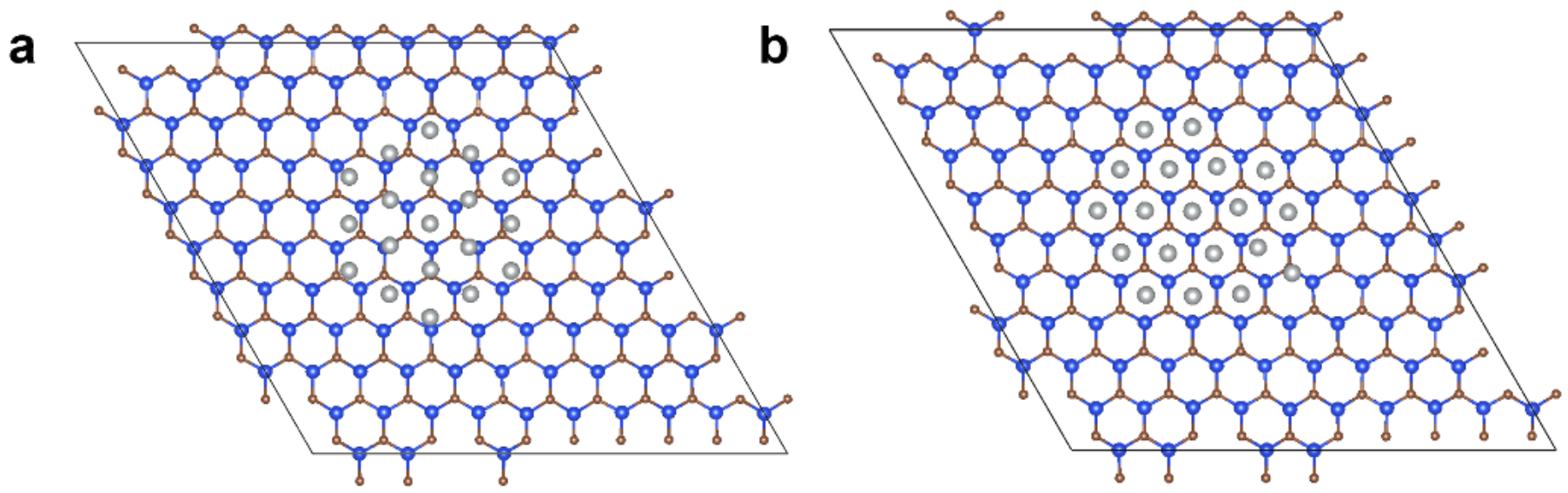

**Fig. S10. Silver cluster calculation**. **(a)** Initial configuration of a finite $Ag_{(2)}$ phase cluster with a diameter of ~12 Å placed on SiC, with only the topmost atomic layer of SiC visualized. **(b)** Upon structural relaxation, the Ag atoms reorganize into $Ag_{(1)}$(hollow) registry, indicating that small $Ag_{(2)}$ clusters are unstable and tend to relax into the $Ag_{(1)}$ phase. This calculation highlights the role of size and registry in phase stabilization of 2D Ag.

## Low-energy electron diffraction (LEED) pattern analysis

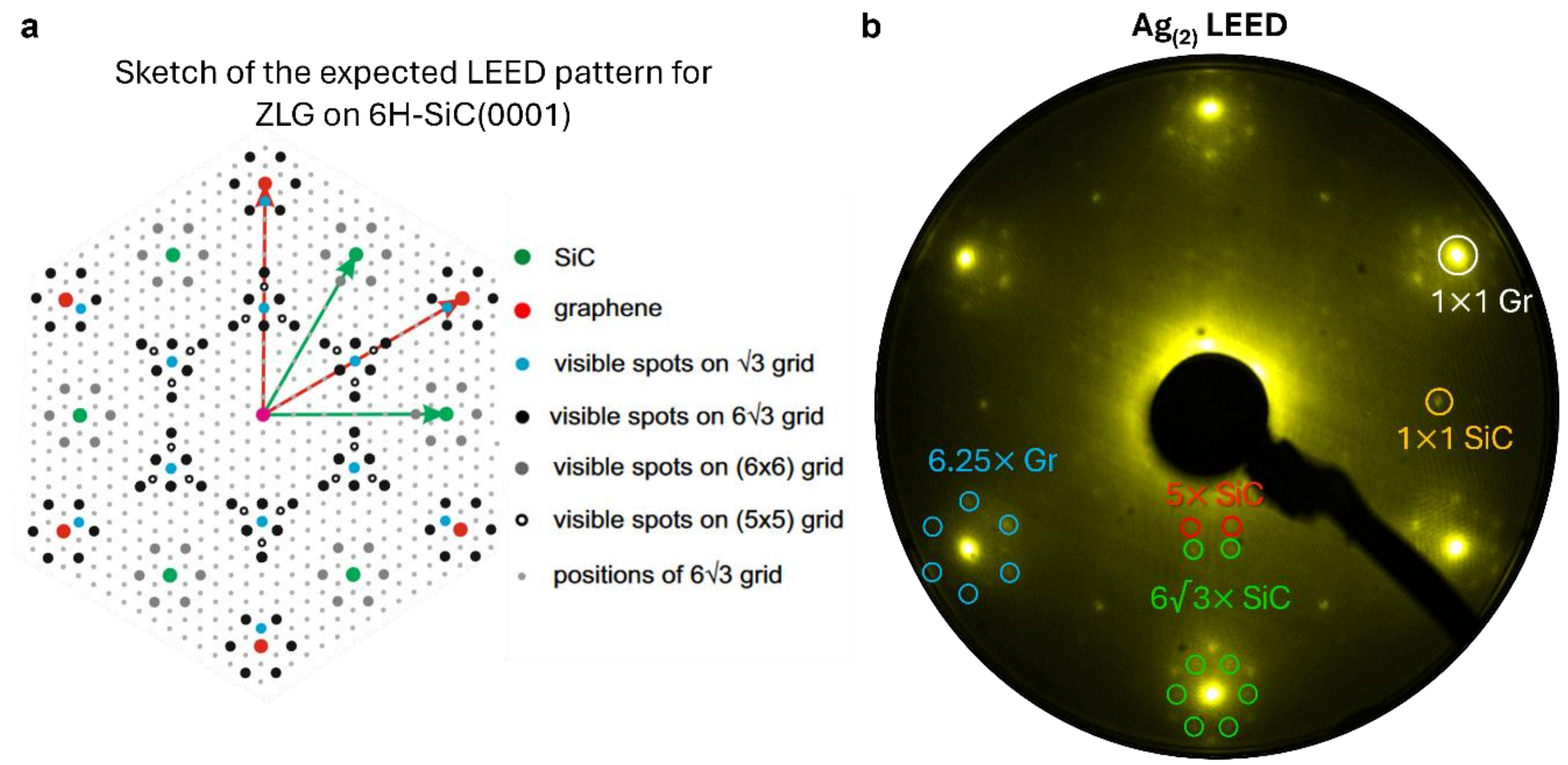

**Fig. S11. LEED pattern analysis for $Ag_{(2)}$.** (a) Simulated LEED pattern for ZLG on 6H-SiC(0001) showing characteristic diffraction spots associated with the SiC substrate and the $(6\sqrt{3} \times 6\sqrt{3})R30°$ buffer layer reconstruction. (b) Experimentally acquired LEED pattern for a predominant $Ag_{(2)}$ sample, where additional satellite spots emerge beyond the standard ZLG pattern. These $Ag_{(2)}$-specific spots can be attributed to the (27:25) commensurate superstructure, confirming long-range ordering in the intercalated Ag layer.

## Table S1: Lattice constants comparison for LEED

| | | | | |
|---|---|---|---|---|
| **Graphene** | 2.46 | 6.25× = 15.375 | | 13× = 32.0 |
| **Silver ($Ag_{(2)}$)** | ~2.98 | $3\sqrt{3}$×(2.96) = 15.375 | $3\sqrt{3}$×(2.98) = 15.5 | 11×(2.93) = 32.2 |
| **SiC** | 3.10 | | 5× = 15.5 | $6\sqrt{3}$× = 32.2 |

**Tab. S1. Lattice constants comparison for LEED**. All values are in angstroms (Å). By analyzing lattice constant mismatches, we determine the in-plane lattice parameter of the $Ag_{(2)}$ phase to be approximately 2.98 Å. This leads to a commensurate 27:25 matching with the SiC substrate—i.e., a $(3\sqrt{3} \times 3\sqrt{3})R30°$ $Ag_{(2)}$ lattice on top of a 5 × 5 SiC supercell.

**Energy-dispersive X-Ray Spectroscopy (EDS) and Transmission Electron Microscopy (TEM) characterization**

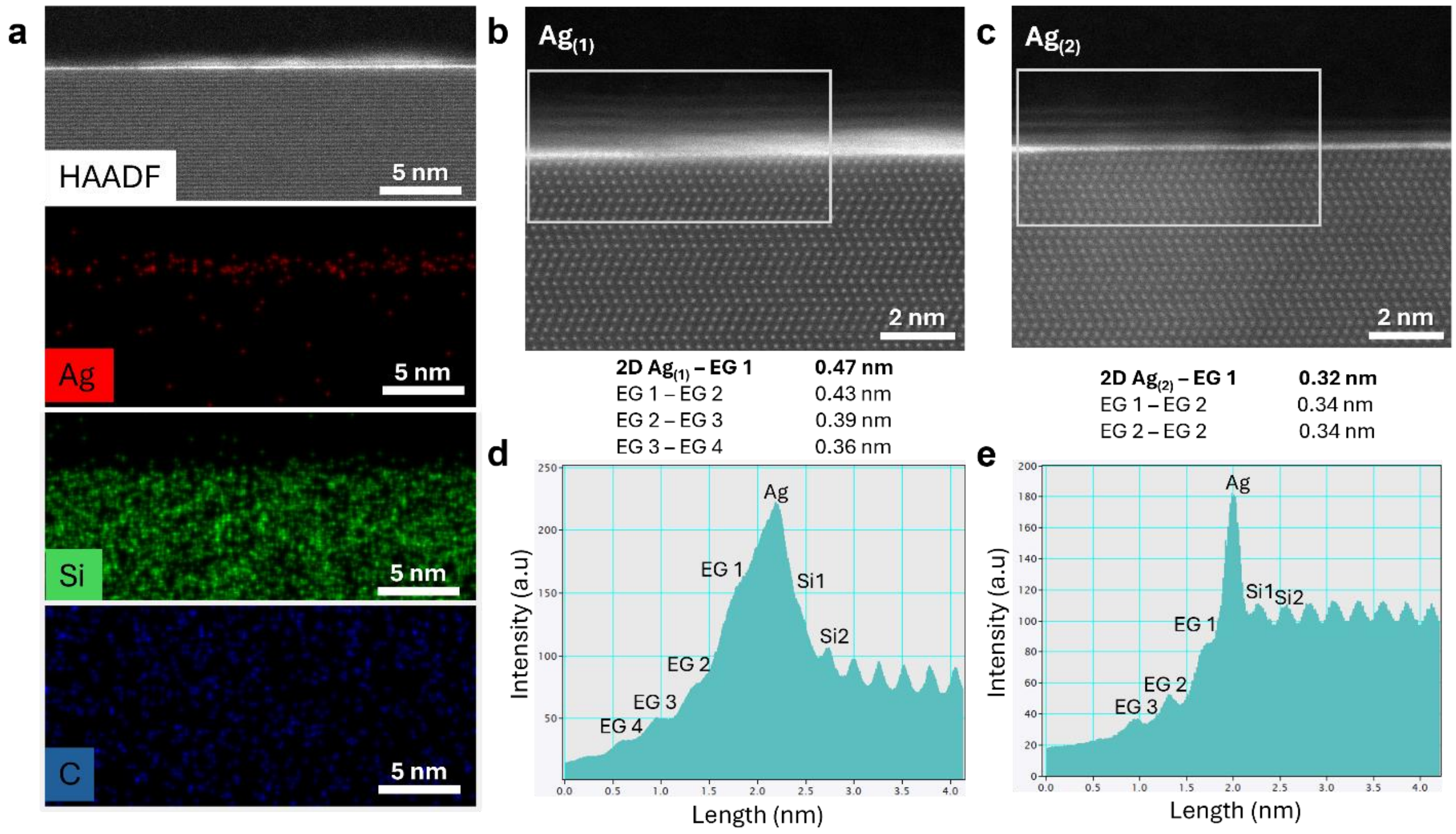

**Fig. S12. EDS and additional TEM characterization of intercalated 2D Ag**. **(a)** High-angle annular dark-field scanning transmission electron microscopy (HAADF-STEM) image and corresponding energy-dispersive X-ray spectroscopy (EDS) elemental maps for 2D $Ag_{(2)}$, highlighting the spatial distribution of Ag, Si and C within the cross-section. The results confirm successful silver intercalation beneath epitaxial graphene, forming a predominantly monolayer Ag structure capped by a single graphene layer. Excess carbon at the top comes from the protective amorphous carbon cap on EG/Ag/SiC. **(b–c)** Cross-sectional HAADF-STEM images of monolayer $Ag_{(1)}$ and $Ag_{(2)}$, respectively, acquired near SiC step edges. **(d–e)** Intensity profiles extracted from the white rectangles in panels **(b)** and **(c)**, respectively, reveal a larger vertical separation between Ag and EG in $Ag_{(1)}$ (0.47 nm) compared to $Ag_{(2)}$ (0.32 nm). Note: Cross-sectional TEM measurements were performed at step edges containing multilayer graphene for better contrast, and may not fully represent terrace-level Ag–EG separation.

**Semiconducting Behavior of $Ag_{(2)}$ in the Presence of Graphene Cap**

First-principles calculations indicate that $Ag_{(2)}$, in its 27:25 commensurate structure with SiC, is intrinsically metallic when graphene is not present due to self-doping from the two extra Ag atoms per supercell, which contribute excess electrons and raise the Fermi level to the conduction band, as shown in the $Ag_{(2)}$ bandstructure in **Fig. S13**. However, experimental observations from angle-resolved photoemission spectroscopy (ARPES, **Fig. S17**) and spectroscopic imaging ellipsometry (SIE) measurements (**Fig. S18**) consistently show that the Ag–SiC system exhibits semiconducting behavior, with the Fermi level lying within the gap. This discrepancy can be reconciled by considering the role of the graphene overlayer. Since the Fermi level of metallic $Ag_{(2)}$ lies above that of freestanding graphene, electron transfer occurs from the Ag–SiC system to the proximate graphene cap, partially depleting the excess charge.

Despite this transfer, the density of states near graphene's Dirac point is small, limiting the extent to which it can accommodate the transferred electrons. As a result, graphene alone cannot fully compensate for the self-doping of $Ag_{(2)}$. Instead, additional charge depletion likely arises from trap states introduced by native point defects in the Ag lattice observed in scanning tunneling microscopy (STM) topography images (**Fig. 3e**). These defects serve as electron sinks, enabling the Fermi level to settle within the bandgap, and therefore rendering $Ag_{(2)}$ semiconducting under realistic experimental conditions.

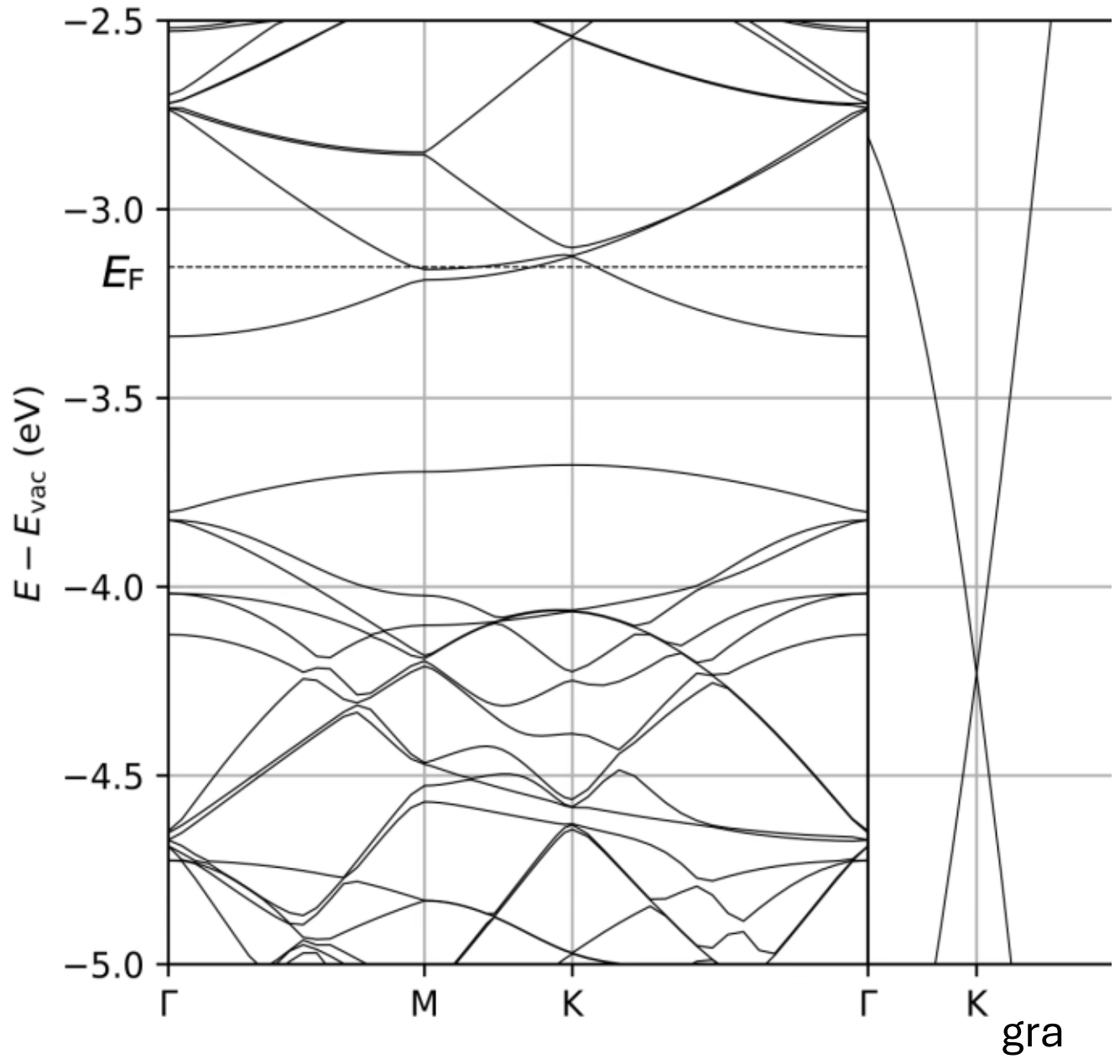

**Fig. S13. Calculated band structure for $Ag_{(2)}$ and comparison with freestanding graphene.** Left: Electronic band structure of the $Ag_{(2)}$ (27:25) phase, referenced to the vacuum level. The Fermi level (dashed line) lies within the conduction band due to self-doping introduced by the two excess Ag atoms per supercell, indicating intrinsic metallicity in the absence of graphene. Right: Dirac cone band structure of freestanding monolayer graphene. The energy offset highlights that the Fermi level of $Ag_{(2)}$ resides above that of graphene, enabling charge transfer when the two systems are brought into proximity. This interaction results in partial electron depletion from the Ag–SiC system into the graphene cap.

**Correlative Characterization of 2D Ag phases through Raman, SEM, AFM and AES mapping**

Correlative multimodal imaging techniques enable precise identification and differentiation of $Ag_{(1)}$ and $Ag_{(2)}$ phases in 2D silver heterostructures. Prior studies have established that $Ag_{(1)}$ and $Ag_{(2)}$ exhibit distinct low-frequency Raman responses and different Auger electron emission intensities due to their differing silver concentrations.[7] Additionally, theoretical and experimental works have proposed that 2D Ag can form various commensurate structures, including 1:1 and 4:3 Ag:Si phases, with 1-4 layer Ag coverage.[8,9] To confirm phase assignment, thickness uniformity, and packing density in our intercalated samples, we employ a combination of Raman spectroscopy, secondary electron SEM, AFM, and AES mapping.

We perform all characterizations on a single terrace of a large-area, phase-segregated sample with optically visible $Ag_{(1)}$ and $Ag_{(2)}$ domains (**Fig. S14a**), thus minimizing contrast artifacts from graphene thickness variation. Raman mapping in the ultra-low frequency (ULF) range confirms the presence of intercalated silver across the terrace, with red areas corresponding to $Ag_{(1)}$ and blue areas to $Ag_{(2)}$. The 2D peak width map clearly reveals uniform bilayer graphene on the terrace and thicker graphene on the edges.[10] After background subtraction, the graphene G and 2D peak positions shift differently depending on the underlying Ag phase—$Ag_{(2)}$ shows a G-peak blueshift and 2D-peak redshift compared to $Ag_{(1)}$, indicating variations in local strain and doping.[11] Decomposition of strain and doping contributions reveals that graphene above $Ag_{(2)}$ experiences greater tensile strain (~0.14%) and higher n-type doping (~$3.2\times10^{13}$ $cm^{-2}$) than graphene above $Ag_{(1)}$ (~0.03%, ~$2.0\times10^{13}$ $cm^{-2}$), supporting stronger interaction and electron transfer at the $Ag_{(2)}$ interface—consistent with calculations in **Fig. S13**, and experimental ARPES (**Fig. S17**) and TEM (**Fig. S12**) measurements .

In secondary electron SEM images acquired on the same terrace, $Ag_{(2)}$ domains appear darker than $Ag_{(1)}$ (**Fig. S14c**). This difference arises from the lower secondary electron emission yield of more n-doped graphene over $Ag_{(2)}$ than $Ag_{(1)}$, which reduces surface band bending and emission efficiency.[12] AFM topography and adhesion mapping further support these findings. Adhesion contrast tracks well with SEM and Raman features, but height maps reveal no measurable step between $Ag_{(1)}$ and $Ag_{(2)}$, confirming that both phases exhibit similar layer thicknesses. If a layer number difference existed, a ~0.3 nm height step—typical for additional silver layers—would have appeared.

To verify the difference in silver packing density between the two phases, we perform Auger Electron Spectroscopy (AES) mapping in the same region (**Fig. S15**). AES intensity maps show $Ag_{(2)}$ domains as brighter than $Ag_{(1)}$, consistent with higher silver content. Since both regions have the same layer thickness and identical bilayer graphene capping, the contrast in AES must arise from differences in Ag atom packing. ImageJ analysis of regions of interest on the terrace yields an $Ag_{(2)}/Ag_{(1)}$ AES intensity ratio of $1.09 \pm 0.01$, aligning closely with the expected 27:25 Ag:Si commensuration ratio (1.08). This observation confirms that $Ag_{(2)}$ adopts a denser packing structure than $Ag_{(1)}$ and further supports our assignment of the $Ag_{(2)}$ phase as the 27:25 commensurate structure on SiC.

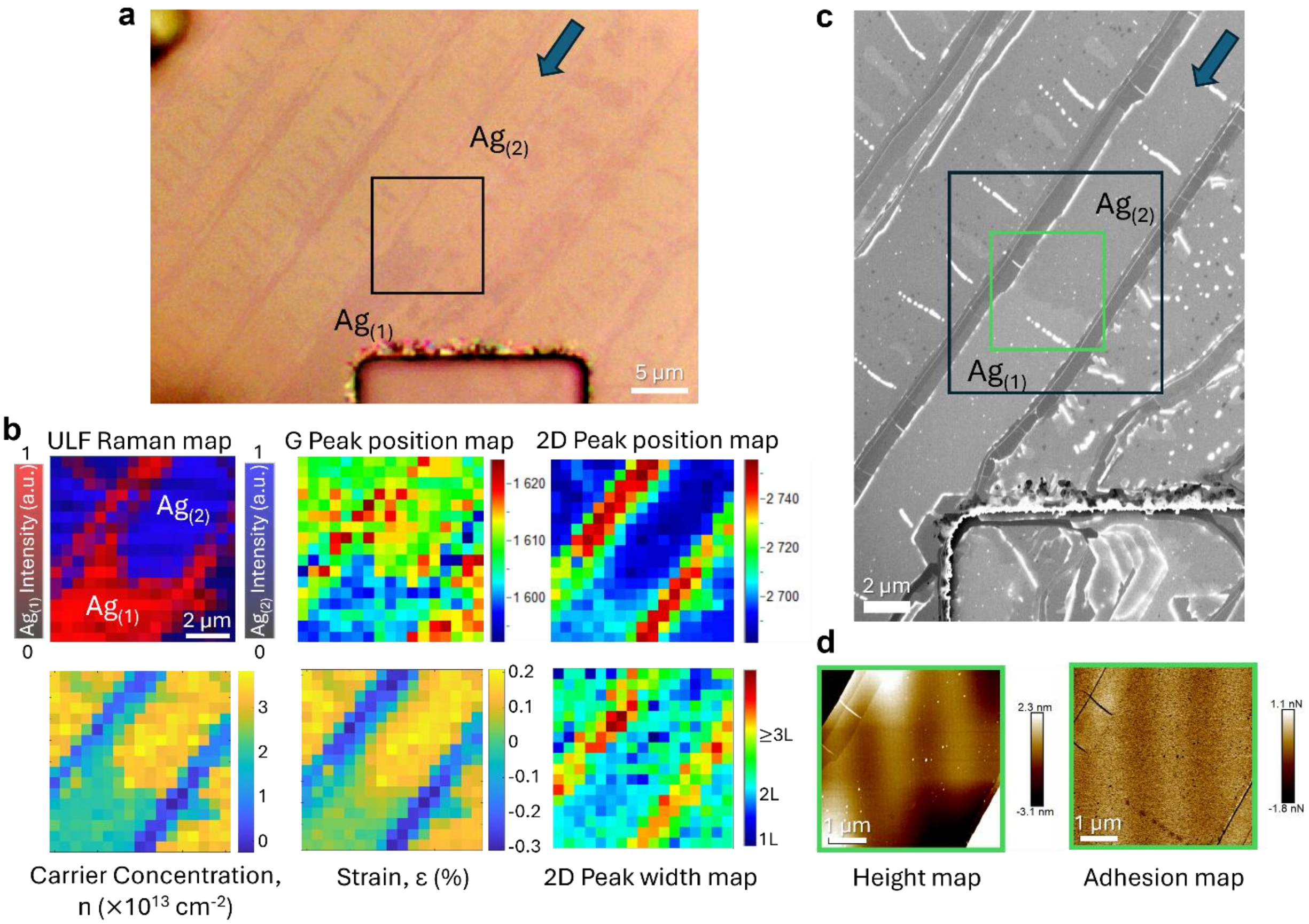

**Fig. S14. Correlative Optical, Raman, SEM and AFM characterization of 2D Ag phases**. **(a)** Optical micrograph of a mixed-phase 2D Ag sample, showing $Ag_{(1)}$ (dark) and $Ag_{(2)}$ (bright) domains on a single terrace (marked with a blue arrow). A lithographic marker at the bottom aids spatial correlation. **(b)** Raman maps from the 10 × 10 μm² region (black square in **(a)**), showing ultra-low frequency (ULF) Raman intensity to distinguish $Ag_{(1)}$ and $Ag_{(2)}$, G and 2D peak positions, 2D peak width (layer number), and extracted strain and carrier concentration maps based on vector decomposition of the G/2D positions. **(c)** Secondary electron SEM image of the same region, showing contrast inversion between phases ($Ag_{(1)}$ – bright, $Ag_{(2)}$ – dark). The blue arrow and black square denote the same terrace and Raman-mapped region, respectively. **(d)** AFM topography and adhesion maps of a 5 × 5 μm² region (green square in **(c)**), across the $Ag_{(1)}/Ag_{(2)}$ boundary, showing no height difference but distinct adhesion contrast correlated with surface features.

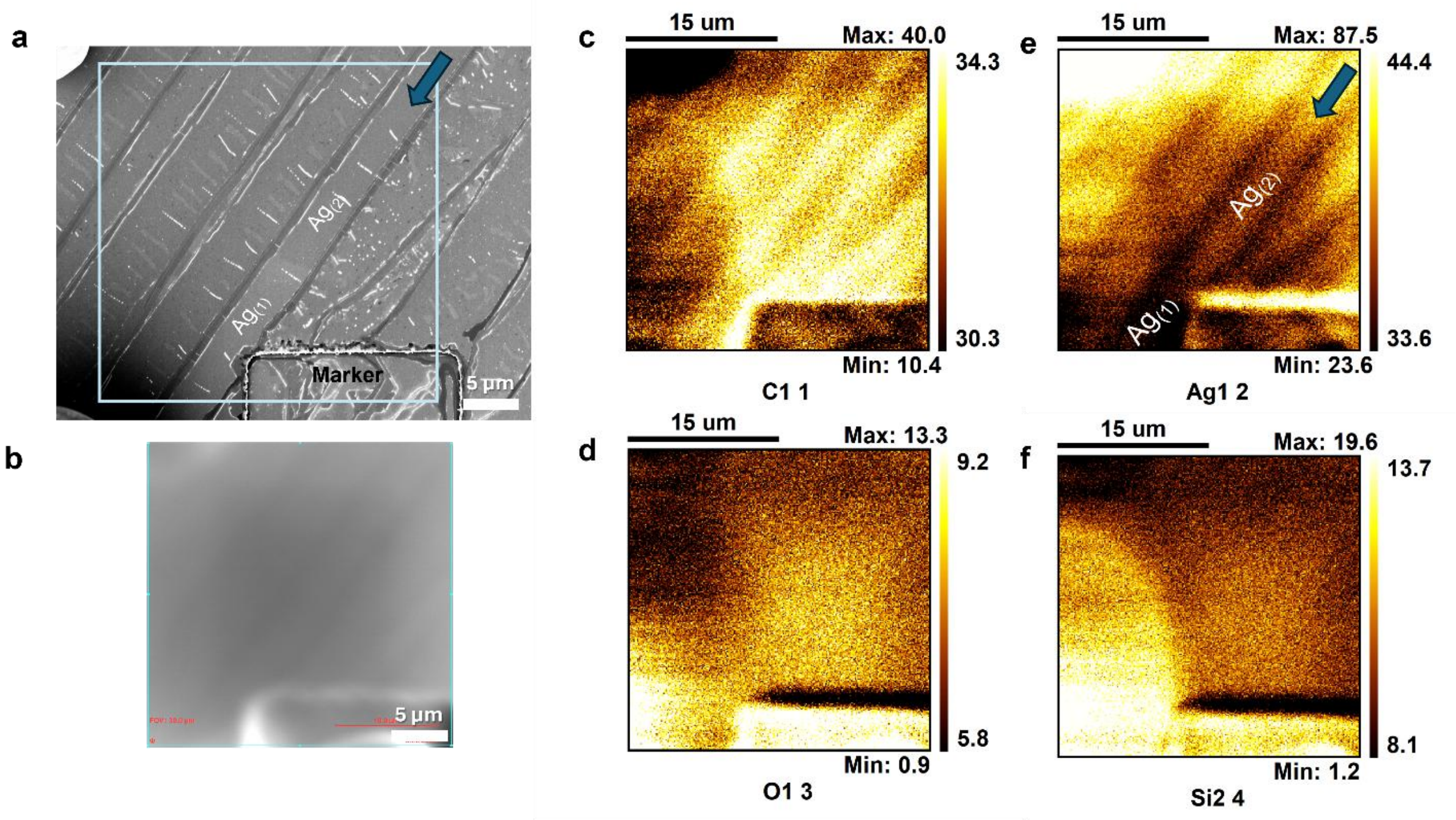

**Fig. S15. Correlative Auger Electron Spectroscopy (AES) and SEM characterization of 2D Ag phases**. **(a)** Secondary electron SEM micrograph of the same sample in **Fig. S14a**, highlighting $Ag_{(1)}$ and $Ag_{(2)}$ regions on a single terrace (blue arrow). The lithographic marker and a silver particle (white spot, top-left) serve as identifiers. **(b)** SEM image from the AES instrument showing the same area. **(c–f)** AES elemental maps from the 30 × 30 µm² region marked with a light blue square in **(a)**, showing distributions of **(c)** carbon, **(d)** oxygen, **(e)** silver, and **(f)** silicon. The blue arrow in **(e)** marks the same terrace as in **Fig. S14**. $Ag_{(1)}$ and $Ag_{(2)}$ regions are labeled.

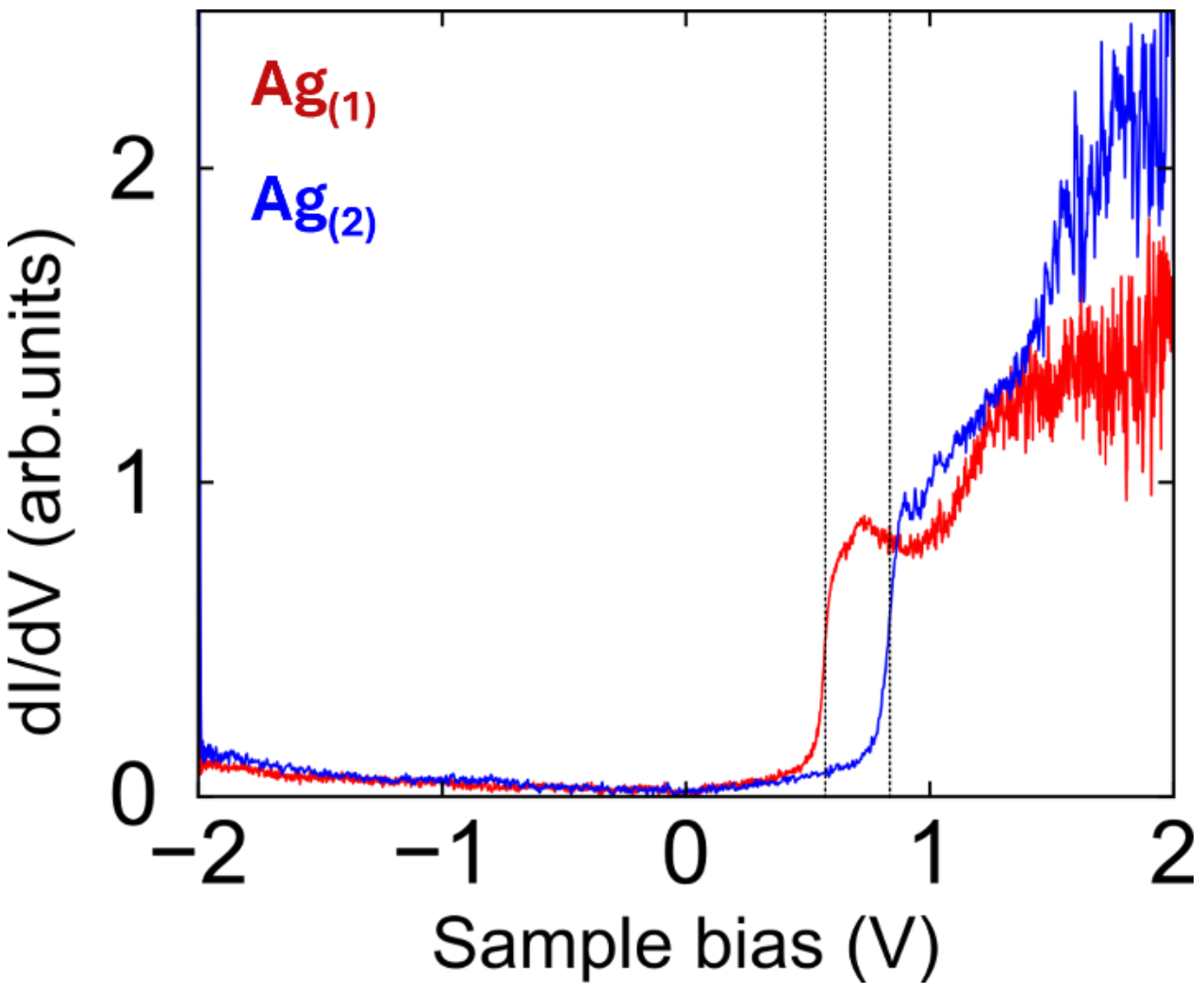

**Fig. S16. Scanning Tunneling Spectroscopy (STS) comparison of 2D Ag phases**. Scanning tunneling spectroscopy (dI/dV) spectra acquired from $Ag_{(1)}$ and $Ag_{(2)}$ regions shown in **Fig. 3(e)** reveal distinct phase-dependent electronic signatures.

## Additional ARPES data and analysis

**Fig. S17a** and **S17b** present the electronic band structures of the $Ag_{(1)}$ and $Ag_{(2)}$-intercalated samples, respectively, along the $\overline{M_{SiC}\Gamma K_{SiC}}$ direction, measured with lab-based 21.22 eV photons. The Brillouin zone (BZ) SiC, which coincides with $Ag_{(1)}$ BZ, is rotated by 30° relative to the graphene BZ as shown in **Fig. S17c**. For both $Ag_{(1)}$ and $Ag_{(2)}$ ARPES data, the most intense features are the graphene $\pi$-band around 3.0 eV binding energy (BE) at $\overline{M}_{gr}$ ($k_{\parallel} = 1.474$ Å$^{-1}$) and the SiC valence band below 2.0 eV BE at Γ. All other observed bands originate from the intercalated Ag layer.[6,13,14] For $Ag_{(1)}$, the valence band maximum (VBM) is located at the $\overline{K}_{SiC}$ point at approximately 0.4 eV BE, and the saddle point at $\overline{M}_{SiC}$ lies near 1.2 eV BE (both are shown by the brown arrows in **Fig. S17a**). In the $Ag_{(2)}$ sample, the VBM is too weak to clearly identify the band maximum, but the saddle point (shown by the blue arrow in **Fig. S17b**) is shifted by approximately 0.2 eV toward higher binding energies.

**Fig. S17(d)** presents the high-resolution Dirac cone for the $Ag_{(2)}$-intercalated sample, measured with 180 eV photon energy. Two distinct Dirac cones are clearly resolved, with Dirac points at approximately 0.3 eV ($E'_{D2}$) and 0.8 eV ($E_{D2}$). The momentum distribution curves (MDCs) (red dots) taken at the Fermi level ($E_F$) of **Fig. 3h** and **Fig. S17(d)** respectively for $Ag_{(1)}$ (**Fig. S17 e(i)**) and $Ag_{(2)}$ (**Fig. S17 e(ii)**) clearly reveal the opposite intensity patterns for QFMLG and QFBLG Dirac cones in these samples. From the fitted data (black line in the figure), the peak separation between the two branches of the Dirac cone corresponding to both QFMLG and QFBLG layer is more than 0.01 Å$^{-1}$ larger for $Ag_2$. In other words, the Fermi pockets of both the QFMLG and QFBLG $\pi^*$ bands are larger in the $Ag_2$ sample. It is important to note that the separation between the $Ag_2$-QFMLG $\pi^*$band peaks in **Fig. S17 e(ii)** correspond to $k_F$= 0.274 Å$^{-1}$, which was used to calculate the charge carrier density in the main text.

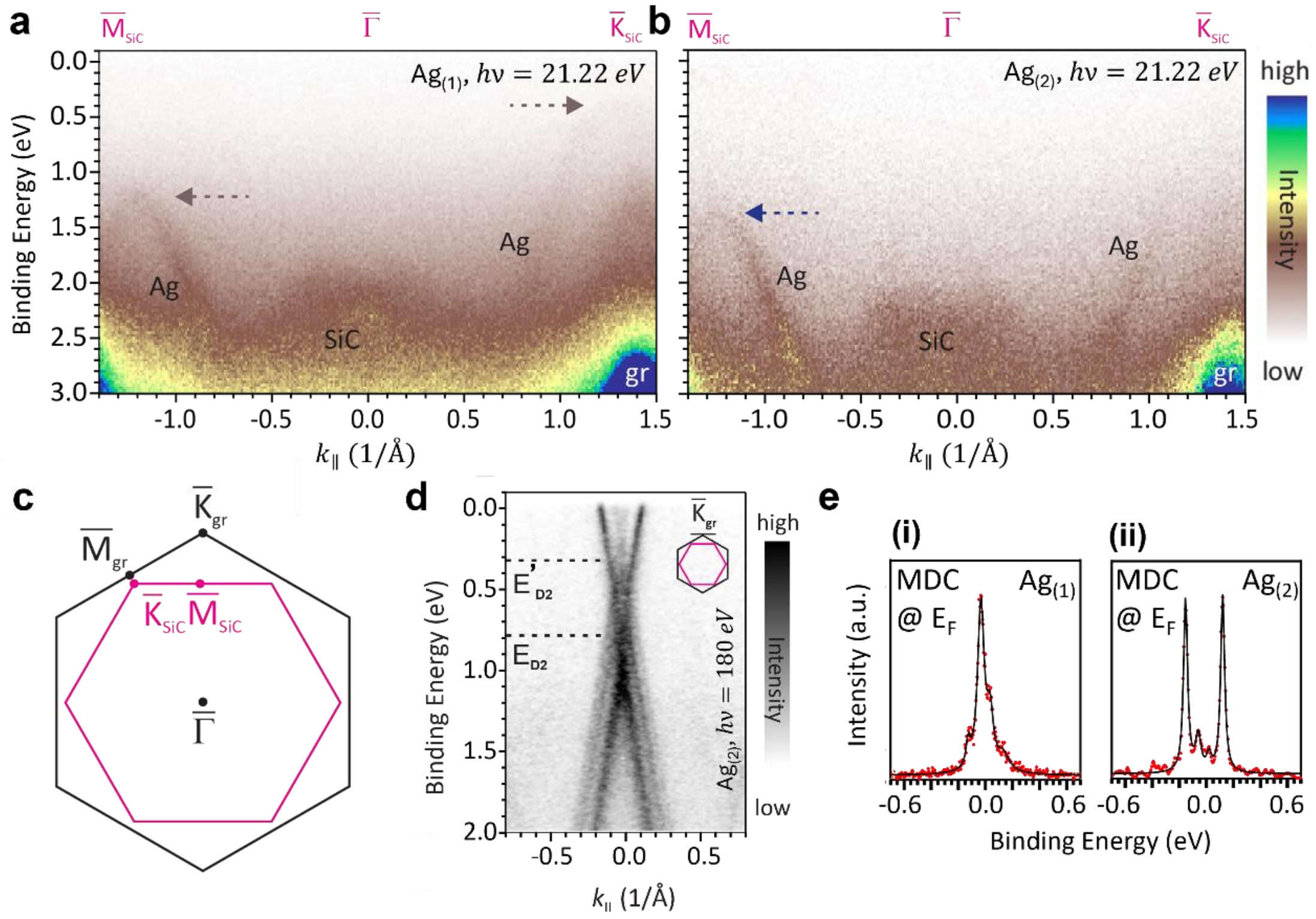

**Fig. S17. Electronic band structure of $Ag_{(1)}$ and $Ag_{(2)}$ intercalated graphene.** Lab-based He Iα measurements of Ag band dispersion along $\overline{M_{SiC}\Gamma K_{SiC}}$ for **(a)** $Ag_{(1)}$ and **(b)** $Ag_{(2)}$. Brown arrows in **(a)** and blue arrows in **(b)** mark characteristic bands of $Ag_{(1)}$ and $Ag_{(2)}$, respectively. **(c)** Brillouin zone (BZ) of Ag/SiC (magenta), rotated and enclosed within the graphene BZ (black). **(d)** High-resolution Dirac cone of $Ag_{(2)}$–QFMLG (180 eV photons), with the corresponding path shown in the inset. **(e)** Momentum distribution curves (MDCs) at the Fermi level ($E_F$) for $Ag_{(1)}$ (**Fig. 3h**) and $Ag_{(2)}$ **(Fig. 3i)** are displayed; red dots indicate raw data, and black lines are fits.

**Determination of the dielectric function from spectroscopic imaging ellipsometry (SIE)**

**Fig. S18a** and **b** show SIE Ψ-maps taken at 450 nm (~2.76 eV) on the $Ag_{(1)}$ (sample A) and $Ag_{(2)}$ (sample B) dominant samples, respectively. Several regions of interest (ROIs) of homogeneous response were chosen from both samples (outlined rectangles, **Fig. S18a** and **b**. The $Ag_{(1)}$ regions on the $Ag_{(2)}$ dominant sample were deemed to small and therefore noisy for further investigation. The hyperspectral maps are manually corrected for small sample movements between measurements taken with different wavelengths in order analyze the response from the identical ROI on the samples for the whole spectral range. . Averaging the response over each ROI results in a set of ellipsometric spectra Ψ(λ), Δ(λ).

In ellipsometry, the change of the polarization state of light being reflected from the surface of a thin multilayer system is measured. Here we use the Jones matrix representation allowing to express the change ρ in the polarization by the ellipsometric angles Ψ(λ), Δ(λ) in the following way:

$$\rho = \frac{r_p}{r_s} = \frac{|r_p|}{|r_s|} e^{i(\delta_{rp} - \delta_{rs})} = \tan \Psi e^{i\Delta} \quad (2)$$

Hereby, $r_p$ and $r_s$ are the complex amplitude reflection coefficients for p- and s-polarized light, which can be rewritten in exponential form with an amplitude $|r_i|$ and a phase shift $\delta_{ri}$. As can be seen, tan(Ψ) can be understood as the amplitude ratio, while Δ is the relative phase shift between p- and s-polarized light reflected from the sample.

The spectra Ψ(λ), Δ(λ) corresponding to sample A can be seen for $Ag_{(1)}$ **Fig. S18d, e** and for $Ag_{(2)}$ in **Fig. S18f** and **g**. The spectra show a consistent response of the polarization change of the reflected light across the measured area, with the spread on the order of the measurement uncertainty.

To extract the complex dielectric function $\varepsilon(\lambda) = \varepsilon_1(\lambda) + i\varepsilon_2(\lambda)$ for $Ag_{(1)}$ and $Ag_{(2)}$ from the measured spectra Ψ(λ), Δ(λ), a comprehensive optical model is constructed and fit to the spectra using regression analysis. The model consists of a stack of flat layers, see **Fig. S18c**, where each layer is fully described by its dielectric function ε(λ) and its thickness t. The top and bottom layers (Air and SiC) are assumed to be semi-infinite. The dielectric function of the 6H-SiC substrate in the visible is well known and can be approximated in the investigated spectral range by a Sellmeier parametrization.[15]

On the top of the structure a thin adsorption layer is included since the measurements are taken in ambient. Taking time-series of spectroscopic ellipsometry measurements of stable CHet 2D metal samples, e.g. fully transitioned $Ag_{(2)}$, we find a shift of the ellipsometric response to lower Δ-values over several months, while the Ψ-spectra stay constant. This behavior is consistent with the accumulation of a transparent layer on top of the sample. The precise nature of this adsorption layer remains unclear. However, it can be satisfactorily modeled using a simple water layer with a variable thickness. For the $Ag_{(1)}$ dominant sample A we find $t_W = 2.39\ nm$, while for the, at point of measurement younger, $Ag_{(2)}$ dominant sample B the fit gives $t_W = 0.75\ nm$.

For graphene the dielectric function varies slightly depending on its preparation, layer number, doping and dielectric environment.[16–18] As discussed, in the case of CHet intercalated 2D Ag, the graphene is n-doped by the underlying metal, compare **Fig. 3f**. To account for these additional charge carriers, we base our graphene model on measurements of field-effect doped bilayer graphene,[17] which uses a biaxial model, with an in-plane Lorentz-Drude response, while the out-of-plane component is ε=1. We model the $Ag_{(1)}$ and $Ag_{(2)}$ dominant samples with the assumption of homogeneous quasi-freestanding bilayer and monolayer graphene, respectively and use the effective thicknesses of $t_{eff}(BL) = 0.95$ nm and $t_{eff}(ML) = 0.475$ nm. See **Fig. S18h** and **i** for the fitted dielectric function of the graphene layers.

2D materials provide an additional challenge to ellipsometry. At thicknesses t < <10 nm, the dielectric function $\epsilon$ and thickness of a layer, can in general not be unambiguously evaluated anymore.[19] Instead, the measured signal is proportional to $(\epsilon - 1)t$. For this reason, the 2D Ag phases are modeled with a constant thickness $t = 0.3$ nm, which was estimated from TEM. The dielectric function was found to be dominated by several absorption peaks, which we describe by (Tauc-) Lorentz dispersion models. The Lorentz dispersion model is the prototypical model used to describe the dielectric response of a semiconductor. The dielectric constant $\epsilon_L$ is parameterized as a function of the photon energy $E$:

$$\epsilon_L(E) = \frac{A_L}{E_L^2 - E^2 - i\Gamma_L E} \quad (3)$$

where $A_L$, $E_L$ and $\Gamma_L$ are the oscillator strength, resonance energy and damping constant, respectively.

The Tauc-Lorentz oscillator is a modified Lorentz oscillator model of asymmetric shape for usage near the bandgap of a material.[19] The imaginary part of the dielectric function is defined as:

$$\epsilon_{TL,2}(E) = \begin{cases} 0 & E \leq E_g \\ \frac{\left(E - E_g\right)^2}{E^2} \frac{A_{TL} E_{TL} \Gamma_{\mathrm{TL}} E}{(E_{TL}^2 - E^2)^2 + \Gamma_{TL}^2 E^2} & E > E_g \end{cases} \quad (4)$$

where $E_g$ is the bandgap. The meaning of $A_{TL}$, $E_{TL}$ and $\Gamma_{\mathrm{TL}}$ is consistent with the Lorentz oscillator. The real part of the dielectric response $\epsilon_{TL,1}$ is calculated from the above using the Kramer-Kronig relations.[20] The total dielectric function is simply the sum of all oscillators and the high energy limit $\epsilon_\infty$:

$$\epsilon = \epsilon_\infty + \sum_i \epsilon_i \quad (5)$$

As can be seen in **Fig. S18j** and **k** the dielectric function of the $Ag_{(2)}$ phase extracted from the optical model show some slight variance between the two samples, with a lower dielectric response from sample B. The differences become clearer looking at the parameters tabulated in **Table S2**. While the dominating absorption peak at 2.19 eV is nearly identical, the higher energy, broad absorption peak is less pronounced on sample B. The size of the effect is too large to be explained by uncertainties in the modeling. The observed differences might arise due to the samples having different dominating graphene thicknesses (bilayer EG for $Ag_{(1)}$ predominant vs monolayer EG for $Ag_{(2)}$ predominant samples), which could lead to differing Fermi levels for $Ag_{(2)}$ between the samples.

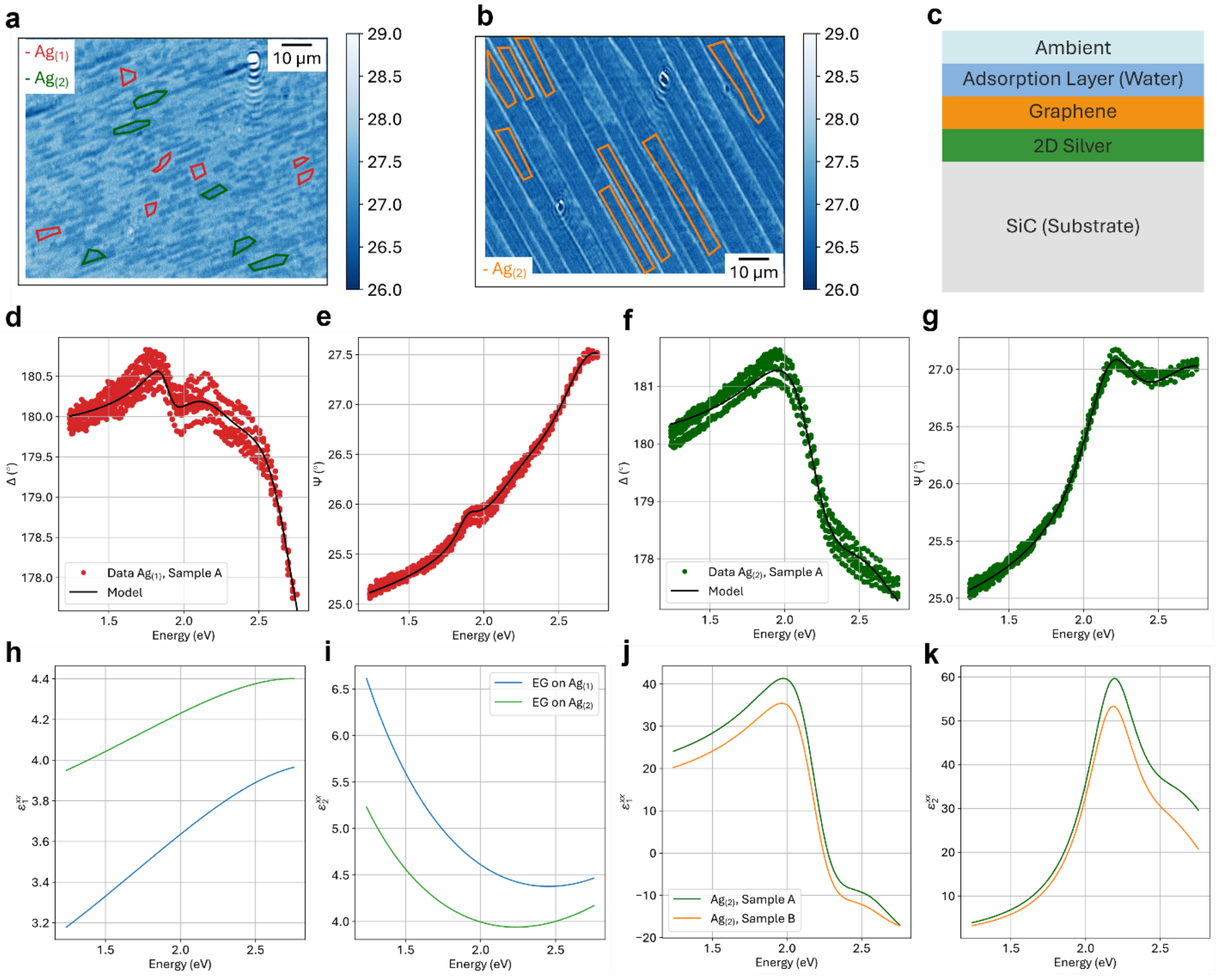

**Fig. S18. Spectroscopic Imaging ellipsometry (SIE) data analysis for Ag$_{(1)}$ and Ag$_{(2)}$.** SIE Ψ-maps taken at 450 nm (~2.76 eV) on the **(a)** Ag$_{(1)}$ (sample A) and **(b)** Ag$_{(2)}$ (sample B) dominant samples, respectively. Regions of interest (ROI) are marked in red (Ag$_{(1)}$), green (Ag$_{(2)}$), and orange (Ag$_{(2)}$). **(c)** Optical model used to extract the dielectric function from SIE data, incorporating the 2D Ag layer, graphene, substrate, and ambient adsorption layer. Corresponding ellipsometric spectra (Ψ(λ), Δ(λ)) for **(d, e)** Ag$_{(1)}$ and **(f, g)** Ag$_{(2)}$, averaged over respective ROIs. **(h)** Real and **(i)** Imaginary parts of the dielectric function ($\varepsilon = \varepsilon_1 + i\,\varepsilon_2$) for EG atop Ag$_{(1)}$ (blue), and Ag$_{(2)}$ (green). **(j)** Real and **(k)** Imaginary parts of the dielectric function for Ag$_{(2)}$, extracted from Sample A (green) and Sample B (orange).

**Table S2: Parameters of the dielectric function of $Ag_{(1)}$ and $Ag_{(2)}$**

| Parameter | sample A, $Ag_{(1)}$ | sample A, $Ag_{(2)}$ | sample B, $Ag_{(2)}$ | Units |
|---|---|---|---|---|
| $\epsilon_\infty$ | 1 | 1 | 1 | - |
| $A_{TL}$ | 11.03797 | - | - | eV |
| $E_{TL}$ | 1.89207 | - | - | eV |
| $\Gamma_{\mathrm{TL}}$ | 0.17293 | - | - | eV |
| $E_g$ | 1.15934 | - | - | eV |
| $E_L$ | 2.28606 | 2.19207 | 2.18533 | eV |
| $A_L$ | 21.87463 | 46.85389 | 43.9079 | $eV^2$ |
| $\Gamma_{\mathrm{L}}$ | 0.65654 | 0.43531 | 0.43865 | eV |
| $E_L$ | 2.7165 | 2.68844 | 2.63125 | eV |
| $A_L$ | 61.52981 | 53.64585 | 33.64682 | $eV^2$ |
| $\Gamma_{\mathrm{L}}$ | 0.52531 | 0.80726 | 0.72899 | eV |

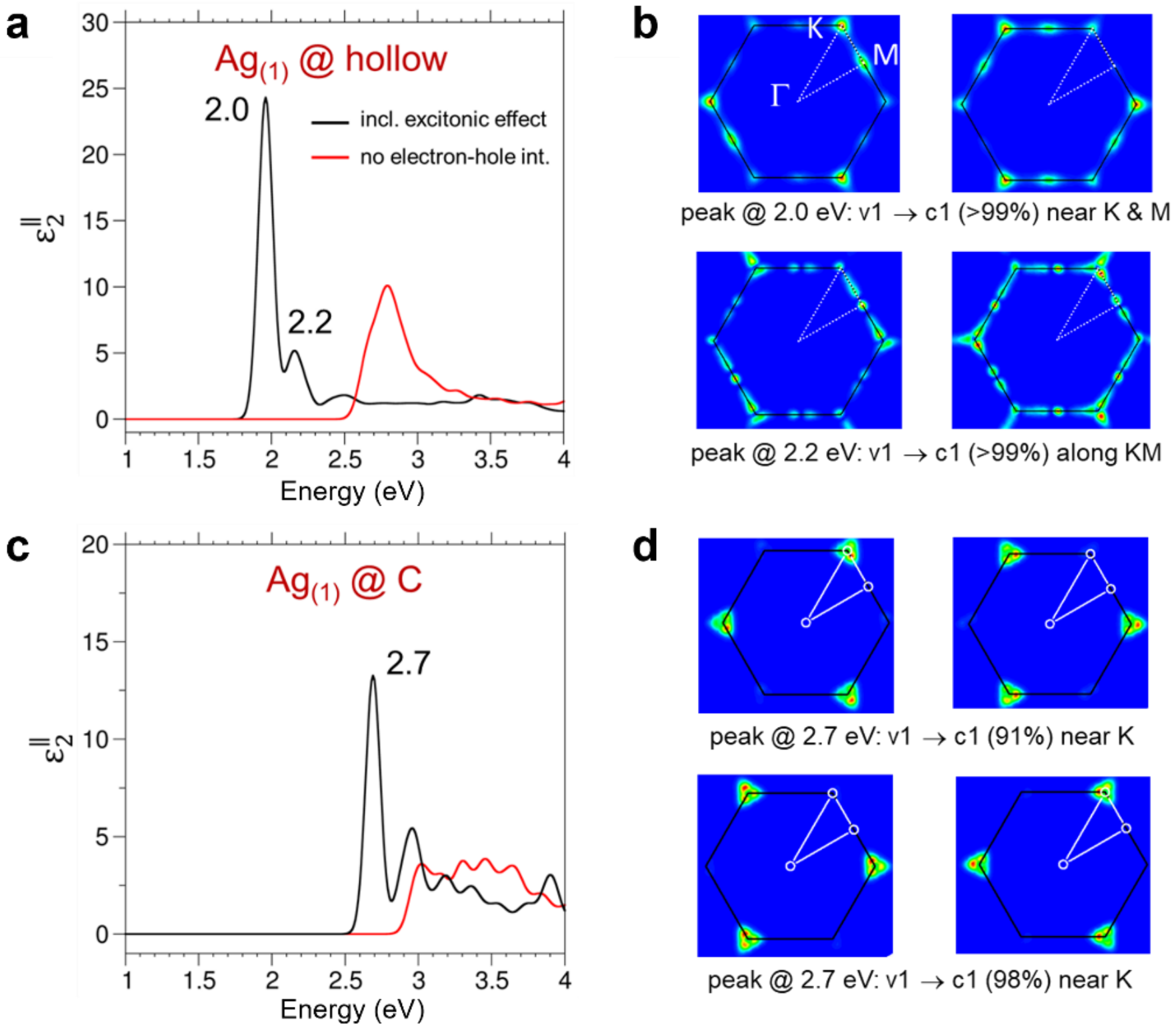

**Fig. S19**. **Optical spectra and k-space distribution of the excitonic transitions for $Ag_{(1)}$.** Computed optical spectra including excitonic effects (GW-BSE) and without electron-hole interactions (GW-IPA) for $Ag_{(1)}$ on (a) hollow site ($Ag_{(1)}$(h)), and (c) carbon site ($Ag_{(1)}$(C)). The k-space distribution transitions leading to (b) excitonic peaks at 2.0 and 2.2 eV for $Ag_{(1)}$(h), and (d) the peak at 2.7 eV for $Ag_{(1)}$(C)). These transitions arise primarily from the topmost valence band to the lowest conduction band.

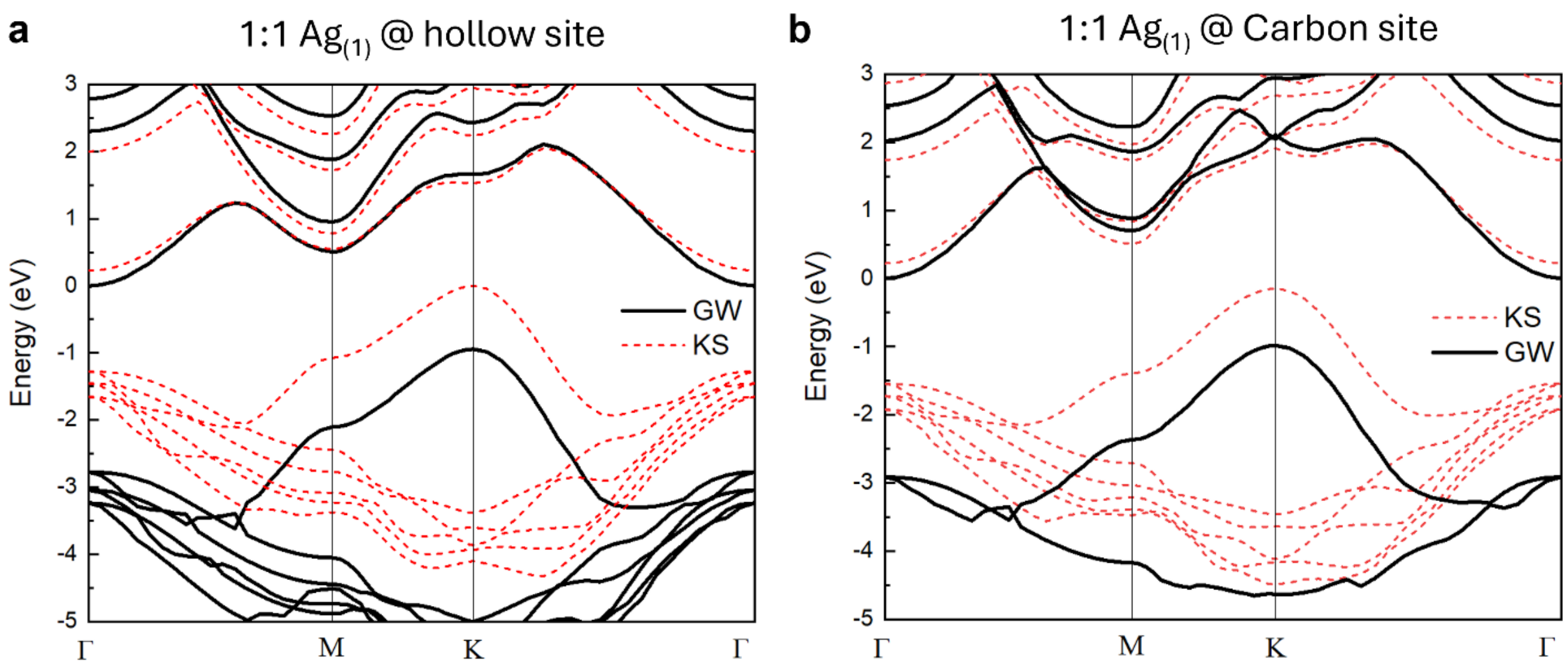

**Fig. S20. Electronic band structure of $Ag_{(1)}$ on different lattice registries.** Electronic band structure for $Ag_{(1)}$ positioned on **(a)** hollow and **(b)** carbon site registries over SiC, calculated using density functional theory (DFT, red dotted lines) and GW (black solid lines) approximations.

**Table S3: Calculated bandgaps for $Ag_{(1)}$ on hollow and carbon sites calculated at DFT and GW level (Fig. S19)**

| $Ag_{(1)}$ lattice site | DFT bandgap (eV) | | GW bandgap (eV) | |
|---|---|---|---|---|
| | Indirect | Direct | Indirect | Direct |
| C | 0.37 | 1.77 (Γ) | 0.98 | 2.91 (Γ) |
| Hollow | 0.23 | 1.50 (Γ) | 0.94 | 2.60 (K) |

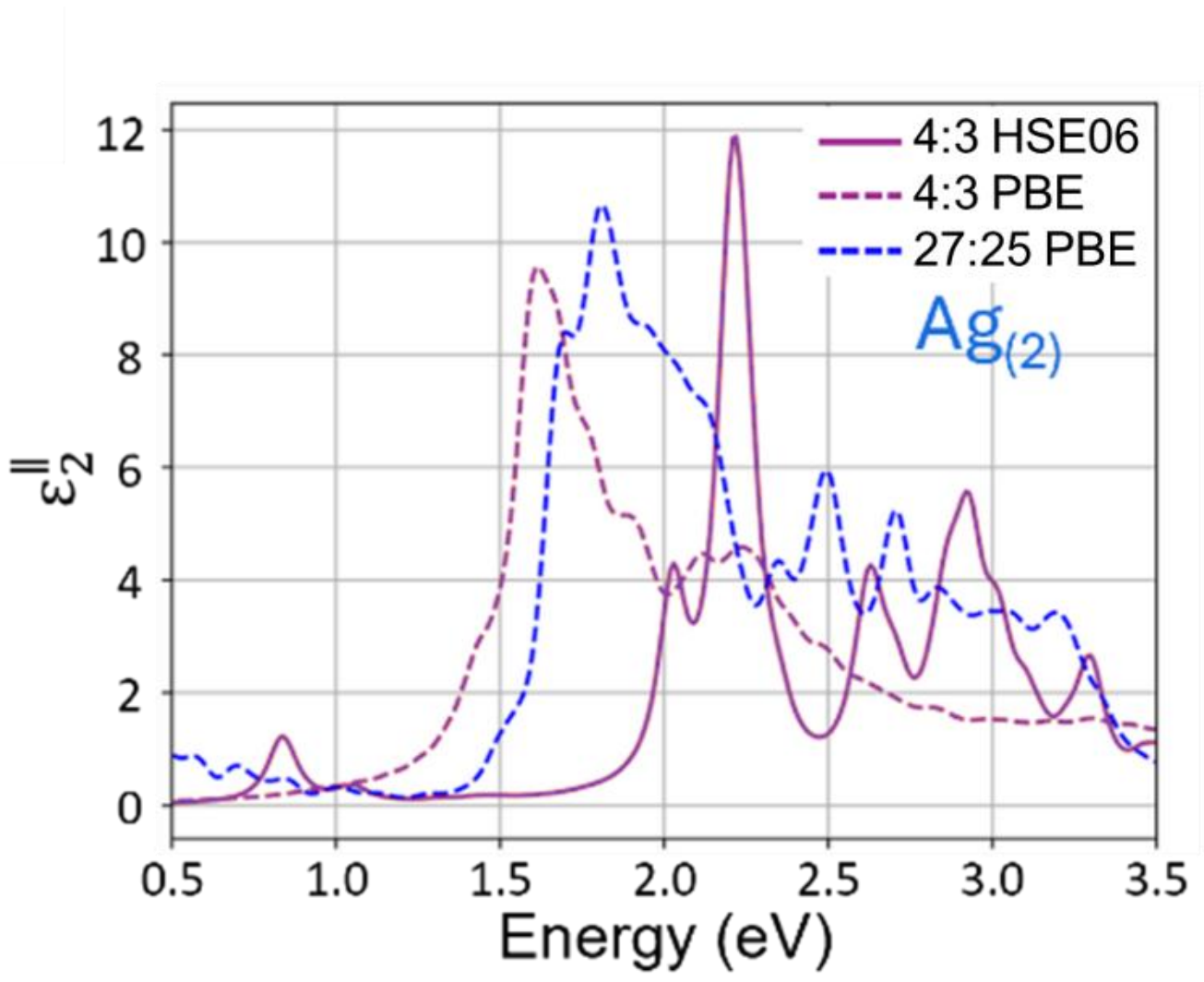

**Fig. S21**. Comparison of $\varepsilon_2$ for $Ag_{(2)}$ using 27:25 and 4:3 Ag:Si models at the PBE level, and 4:3 at the HSE06 level. The spectra calculated at the PBE level using 27:25 and 4:3 Ag:Si models show reasonable agreement.

## $\chi^{(2)}$ Analysis from second harmonic generation (SHG) images

The calculation of $\chi^{(2)}$ was done following methods described by Steves *et al.*[21] The total energy of the detected photons per pixel ($E_v$) of each SHG image is calculated by

$$E_v = \frac{cts}{g} \cdot \frac{S}{QE} \cdot 3.65$$

where cts is the photon count per pixel of the taken SHG image, S is the CCD sensitivity, g is the electron multiplying, QE is the quantum efficiency, and 3.65 is a physical constant accounting for electron generation in Si. In our procedure, S = 4, g = 300, and QE = 0.71. The SHG power is then calculated by $P_{2\omega} = E_v/(t \times accu)$ where the collection time t = 10 s and the number of accumulations accu = 40. The sheet susceptibility $\chi^{(2)}_{sheet}$ ($nm^2/V$) was then calculated following Shen *et al.*[22] and Clark *et al.*[23],

$$|\chi^{(2)}_{sheet}|^2 = \frac{P_{2\omega}(n_\omega^{SiC} + 1)^6 \epsilon_0 c^3 A}{2048\pi^3 \omega^2 P_\omega^2}$$

Where ω is the fundamental frequency, $n^{SiC}_\omega$ is the refractive index of 6H-SiC at the fundamental frequency adopted from literature,[24] $P_\omega$ is the incident laser power at the fundamental frequency, and A is the area illuminated by the exciting pulse. Finally, the effective susceptibility $\chi^{(2)}_{eff}$ (nm/V) is then through the relation $\chi^{(2)}_{eff} = \chi^{(2)}_{sheet}/d$, where d is the Ag monolayer thickness obtained from TEM. Five SHG images were taken for two different $Ag_{(1)}$ samples and two different $Ag_{(2)}$ samples, thus the calculated $\chi^{(2)}$ was averaged over ten ROIs for each Ag phase. The calculated $\chi^{(2)}_{sheet}$ for $Ag_{(1)}$ and $Ag_{(2)}$ are 0.001 ± 0.0001 and 3.2 ± 0.08 $nm^2/V$, respectively. And $\chi^{(2)}_{eff}$ for $Ag_{(1)}$ and $Ag_{(2)}$ are 0.004 ± 0.0005 and 11.4 ± 0.3 nm/V, respectively.

**Supplementary References**